\documentclass[preprint,11pt,authoryear,5p, twocolumn, times]{elsarticle}
\usepackage{amsmath,amssymb,booktabs}
\usepackage{graphicx}
\usepackage{multirow}
\usepackage{hyperref}
\usepackage[table,xcdraw,dvipsnames]{xcolor}

\begin{document}

\begin{frontmatter}

\title{Quantile Regression Postprocessing with Negative Electricity Price Forecasts}

\author[wu]{Bartosz Uniejewski\corref{cor1}}
\ead{bartosz.uniejewski@pwr.edu.pl}
\cortext[cor1]{Corresponding author: Bartosz Uniejewski}

\author[ox]{James W. Taylor}
\ead{james.taylor@sbs.ox.ac.uk}

\address[wu]{Wroc\l aw University of Science and Technology, Poland}
\address[ox]{Sa\"{\i}d Business School, University of Oxford, United Kingdom}

\begin{abstract}

{Negative electricity prices create market-relevant risks for trading, bidding, renewable revenues, and flexibility valuation.} Quantile regression-based postprocessing is a widely used approach for converting point forecasts into predictive distributions. We show that the standard specification, which imposes a single linear relationship between the point forecast and predictive quantiles, can become structurally incoherent when the conditioning point forecast is negative. In such cases, the resulting predictive distribution may exhibit not only ordinary quantile crossing, where adjacent quantiles are in the wrong order, but also a severe version of crossing, which we refer to as $tail$-$quantile$ $inversion$. {This can reverse downside and upside risk signals.} To address this problem, we propose sign-dependent QR-based postprocessing specifications that allow the quantile function to differ between positive and negative point forecasts. {Using German day-ahead electricity prices as the main empirical setting, we show that these specifications improve predictive accuracy, particularly for negative point forecasts and extreme probability levels. When combined with joint estimation across quantiles, they eliminate quantile crossing and tail-quantile inversion. A robustness analysis for the U.S. SPP market confirms the main coherence results, although the gains in predictive accuracy are less pronounced. We demonstrate the economic relevance of the forecasts using a battery arbitrage strategy, where probabilistic forecasts improve trading performance relative to a point-forecast benchmark and the sign-dependent specifications can provide additional gains.} The results demonstrate that the sign of the point forecast should be explicitly accounted for when constructing probabilistic forecasts in markets where negative prices can occur.
\end{abstract}


\end{frontmatter}

\section{Introduction}

{Probabilistic electricity price forecasting plays an increasingly important role in modern electricity markets \citep{now:wer:18,han:bar:25}.} Market participants are exposed not only to expected price levels, but also to the risks associated with extreme outcomes, price spikes, and negative prices \citep{fan:gam:pro:13}. For this reason, predictive distributions, often represented by a set of conditional quantiles, are more informative than point forecasts alone and provide a natural basis for risk-aware decision-making in trading, bidding, hedging, and operational planning \citep{gne:kat:14}. {This distributional perspective is particularly important in electricity
markets, where price risk is shaped by limited short-term storability, variable
renewable generation, and the increasing role of flexible technologies.}

{{The economic need for probabilistic forecasts is particularly evident in negative-price regimes, which have been linked to renewable integration, system flexibility, and changes in electricity price formation \citep{fan:gam:pro:13,nic:10,mac:20,hui:ste:22}.} Although negative prices typically account for only a limited share of hours, they are economically important. For example in Germany, negative prices have direct financial consequences for renewable generators. Under the EEG 2014\footnote{Erneuerbare-Energien-Gesetz (EEG) 2014, Bundesgesetzblatt I, p. 1066, Section 24: reduction of support in periods of negative prices} six-hour rule, eligible renewable generators did not receive support payments for periods in which day-ahead spot prices were negative for at least six consecutive hours. Under the EEG 2021\footnote{Erneuerbare-Energien-Gesetz (EEG) 2021, Bundesgesetzblatt I, p. 3138, Section 51: reduction of support in periods of negative prices}, this threshold was shortened to four consecutive hours. Negative prices therefore affect renewable revenues, bidding incentives, and the value of flexible technologies such as storage and demand response. Accurately predicting this part of the price distribution is thus crucial for market participants. {The relevant forecasting task is therefore not only to predict whether prices will be negative, but also to provide coherent signals about downside and upside price risk in such market states.}}

A common way to obtain probabilistic forecasts is to postprocess existing point forecasts \citep{wal:hen:zie:gne:24}. In this setting, the point forecast is treated as a summary of the current forecasting situation, and the postprocessing model converts it into a predictive distribution or a set of prediction intervals. Several approaches can be used for this purpose, including historical simulation (HS), conformal prediction \citep[CP;][]{rom:pat:can:19}, isotonic distributional regression \citep[IDR;][]{hen:zie:gne:21}, and quantile regression (QR) postprocessing \citep{now:wer:15}. These methods differ in how they use past forecast errors, how strongly they rely on the point forecast, and whether they estimate the full conditional distribution or only selected prediction intervals.

QR postprocessing is particularly attractive in this context because it estimates predictive quantiles directly as functions of the available point forecast. {It is closely related to quantile regression averaging (QRA), where several point forecasts are used as regressors for predictive quantiles \citep{now:wer:15}. In the single-forecast setting considered here, no averaging across competing forecasts is involved, and we therefore use the term \textit{QR postprocessing}.} This approach makes it possible to construct a dense grid of conditional quantiles, for example at probability levels $0.01,0.02,\ldots,0.99$, and hence to approximate the entire predictive distribution.

Representing uncertainty through a dense grid of quantiles is useful not only for forecast evaluation, but also {for market-oriented applications such as bidding, storage operation, and short-term risk management \citep{nit:wer:23,jan:woj:22,nar:zie:22}.} Compared with historical simulation, QR postprocessing can adapt the predictive distribution to the point forecast rather than relying on the unconditional empirical distribution of past forecast errors. Compared with many residual-based conformal prediction procedures, it can more naturally accommodate asymmetric, skewed, and heavy-tailed distributions \citep{koe:hal:01}, while providing direct control over individual parts of the distribution, including the lower and upper tails.

{At the same time, QR postprocessing also raises coherence issues. When quantiles are estimated separately, the resulting predictive distribution need not be monotone in the probability level, leading to quantile crossing \citep{par:mad:aub:kan:gas:wan:22}. This problem can be addressed by non-crossing estimation methods or by ex-post rearrangement of the predicted quantiles. {For electricity-market applications, however, nonmonotone quantile forecasts are not only a statistical inconvenience. Since lower and upper quantiles are often interpreted as indicators of downside and upside price risk, violations of monotonicity can generate conflicting signals for market decisions.} In this paper, we focus on a related but more specific issue: in standard single-forecast QR postprocessing, the same linear relationship between the point forecast and each predictive quantile is imposed regardless of whether the point forecast is positive or negative. We show that this sign-invariant specification becomes misspecified when point forecasts take both positive and negative values.

To address this issue, we propose sign-dependent QR postprocessing specifications that allow the relationship between the point forecast and the predictive quantiles to differ across positive and negative forecast regimes. We consider two alternatives to the standard specification: a sign-decomposed model, which separates the positive and negative components of the point forecast, and a sign-specific model, which estimates different relationships depending on the sign of the forecast. These specifications are combined with both independent quantile estimation and joint estimation of multiple quantiles under monotonicity constraints. {The proposed models therefore preserve the simplicity of QR postprocessing while allowing price-risk behaviour to change across positive and negative forecast regimes.}

{The main contribution of this paper is to show that a widely used probabilistic price-forecasting tool can deliver economically incoherent risk signals in negative-price regimes, and to propose simple sign-dependent specifications that address the underlying regime misspecification.} Using German day-ahead electricity prices as an empirical illustration, we show that these specifications improve predictive performance, particularly for negative point forecasts and distributional tails, and that, when combined with joint quantile estimation, they eliminate severe quantile crossing. {The results link a technical coherence problem in quantile forecasts to an economically relevant failure in the assessment of lower- and upper-tail electricity price risk.}

{To complement the statistical evaluation, we also examine the economic value
of the probabilistic forecasts using a simple day-ahead battery arbitrage
strategy. The results show that incorporating distributional information
improves trading performance relative to a strategy based solely on point
forecasts, while the proposed sign-dependent specifications can provide
additional economic gains.}

{Our empirical comparison is deliberately restricted to QR-based postprocessing specifications. We do not benchmark QR against alternative probabilistic forecasting methods, as such comparisons are already well established in the electricity-price forecasting literature. Recent studies show that QRA remains a strong and competitive benchmark: \citet{lip:uni:wer:24} compare it with conformal prediction, IDR, and distributional neural networks, while \citet{leb:das:pap:sch:26} evaluate QRA alongside a broad range of statistical and deep-learning probabilistic models. In both studies, QRA performs very competitively. In our empirical analysis, we isolate the specific structural problem of linear QR postprocessing considered in this paper and assess whether the proposed sign-dependent modifications resolve it.}

{To assess whether these findings are specific to the German market, we
repeat the empirical analysis for the Southwest Power Pool (SPP)
market in the United States. The robustness results, reported in \ref{sec:spp},
confirm the main findings concerning forecast coherence, although the gains in
predictive accuracy are less pronounced. This comparison suggests that the
severity of the problem depends not only on how frequently negative prices
occur, but also on their magnitude.}

The remainder of the paper is organized as follows. Section~\ref{sec:motivation} motivates the paper by showing how negative point forecasts can lead to incoherent predictive distributions. Section~\ref{sec:data} describes the data, forecasting setup, and point forecasting models used in the empirical study. Section~\ref{sec:method} presents the quantile regression postprocessing specifications, including the proposed sign-dependent variants. Section~\ref{sec:results} reports the empirical results in terms of predictive accuracy, tail performance, and quantile crossing. {Section~\ref{sec:economic} evaluates the economic usefulness of the probabilistic forecasts using a battery arbitrage strategy. Section~\ref{sec:conclusion} summarizes the main findings and discusses their implications for probabilistic electricity price forecasting and market applications.}

\section{Motivation}
\label{sec:motivation}

\begin{figure}[b!]
    \centering
    \includegraphics[width=\linewidth]{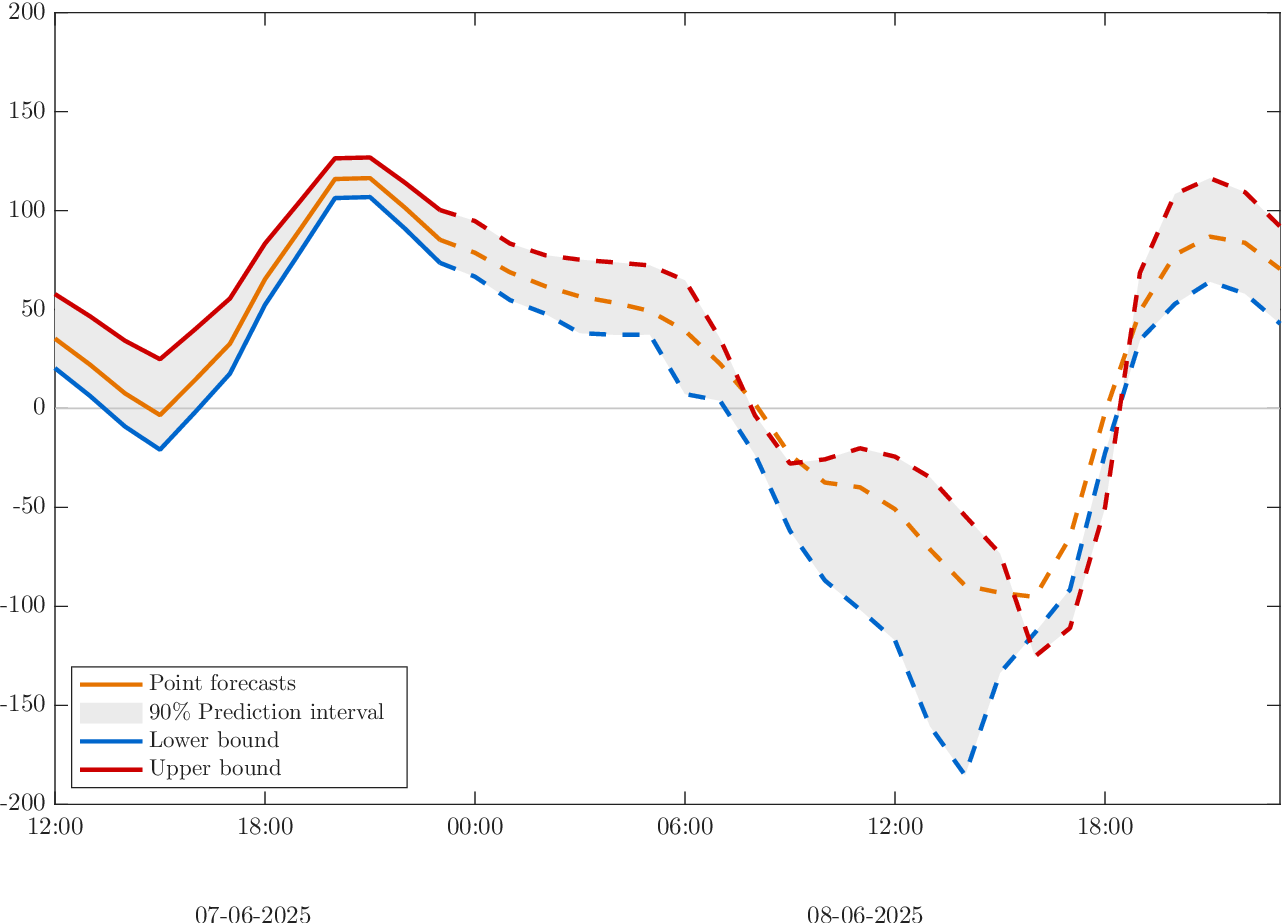}
\caption{{Example of tail-quantile inversion on June 8, 2025, obtained from LEAR point forecasts postprocessed with the SI-I specification using a 182-day calibration window (for details see Section \ref{sec:method}). Blue and red lines denote the 5\% and 95\% quantiles, respectively. Solid lines show the in-sample fit, while dashed lines show the corresponding out-of-sample forecasts for the 24 delivery hours.}}

    \label{fig:motivation}
\end{figure}

The problem considered in this paper is illustrated in Figure~\ref{fig:motivation}. The figure shows a real forecasting instance for 08-06-2025, in the German day-ahead electricity market. Blue and red lines denote the 5\% and 95\% quantiles, respectively. The quantiles were obtained separately for each of the two probability levels by using QR to postprocess a point forecast. Solid lines show the in-sample fitted quantiles, while dashed lines show the corresponding out-of-sample forecasts for the 24 delivery hours of that day. The point forecasts used as conditioning variables are negative for part of the day, and the resulting predicted quantiles no longer preserve the natural ordering of the predictive distribution. This is not a hypothetical numerical artifact, but an outcome obtained from the standard single-forecast QR postprocessing specification.

The source of the problem lies in the standard QR postprocessing of a single point forecast. In its simplest form, for each probability level $\alpha$, the predicted quantile is modeled as a linear function of the same point forecast,
\begin{equation}
    Q_{\alpha}(P_{d,h} \mid \widehat{P}_{d,h})
    =
    a_{\alpha} + b_{\alpha}\widehat{P}_{d,h},
    \label{eq:standard_qr}
\end{equation}

where $\widehat{P}_{d,h}$ denotes the point forecast for day $d$ and delivery hour $h$, and $a_{\alpha}$ and $b_{\alpha}$ are parameters estimated separately for each $\alpha$ using QR. 

{To see how the problem can arise, consider two probability levels ($\alpha_j > \alpha_i$). The difference between the corresponding fitted quantiles is}

\begin{align}
    Q_{\alpha_j}(P_{d,h} \mid \widehat{P}_{d,h})
    & -
    Q_{\alpha_i}(P_{d,h} \mid \widehat{P}_{d,h})
     \nonumber \\ 
     & =
    (a_{\alpha_j}-a_{\alpha_i})
    +
    (b_{\alpha_j}-b_{\alpha_i})\widehat{P}_{d,h}.
    \label{eq:quantile_diff}
\end{align}

{When point forecasts are positive, a natural coefficient pattern is ($b_{\alpha_j} > b_{\alpha_i}$): higher quantiles respond more strongly to an increase in the point forecast than lower quantiles. In that case, the slope component in \eqref{eq:quantile_diff} contributes positively to the distance between the higher and lower quantiles, helping to preserve the ordering of the fitted quantile function.}

The same coefficient pattern becomes problematic when the point forecast is negative. If $\widehat{P}_{d,h}<0$, the term $$(b_{\alpha_j}-b_{\alpha_i})\widehat{P}_{d,h}$$ changes sign. A slope pattern that expands the distance between quantiles for positive point forecasts can therefore reduce, or even reverse, that distance for negative point forecasts. In extreme cases, this leads to
$Q_{\alpha_j}(P_{d,h} \mid \widehat{P}_{d,h}) < Q_{\alpha_i}(P_{d,h} \mid \widehat{P}_{d,h})$
even though $\alpha_j>\alpha_i$.

Figure~\ref{fig:motivation} shows that this issue is more severe than ordinary quantile crossing. In many applications, quantile crossing refers to small local violations, often between neighboring probability levels, which can be viewed as a technical inconvenience and removed by rearranging or sorting the predicted quantiles. In Figure~\ref{fig:motivation}, however, we show that the problem can take the form of a large-scale reversal of the predictive distribution. Even the 5\% and 95\% quantiles may be inverted for some delivery hours, indicating that the lower tail of the predictive distribution is placed above the upper tail.

{This distinction matters because existing approaches to non-crossing quantile estimation primarily aim to enforce or restore monotonicity. These include constrained and nonparametric estimators \citep{bon:rei:wan:10,det:vol:08}, nonlinear quantile-function representations that impose monotonicity directly \citep{can:18,par:mad:aub:kan:gas:wan:22}, and ex-post rearrangement of estimated quantiles \citep{che:f-v:gal:10}. Such methods can produce monotone quantile forecasts, but they do not by themselves address the mechanism described above. The reversal arises from the sign-invariant structure of the standard specification: a coefficient pattern that is coherent for positive point forecasts can become incoherent when the point forecast is negative, and vice versa.}

A purely mechanical correction, such as sorting the predicted quantiles after estimation, can restore monotonicity ex post \citep{che:f-v:gal:10}, but it does not address the underlying misspecification. Rearrangement changes the order of the predicted values, but it does not change the relationship learned by the model between the point forecast and the quantile function. {Moreover, the issue is not confined to the forecasts that are negative at the prediction stage. Since the standard specification is estimated on a calibration sample containing both positive and negative point forecasts, its coefficients represent a compromise across the two regimes. Negative-forecast observations may therefore influence the calibration of the model for positive point forecasts.}

\section{Data and empirical framework}
\label{sec:data}

\begin{figure*}
    \centering
    \includegraphics[width=0.99\linewidth]{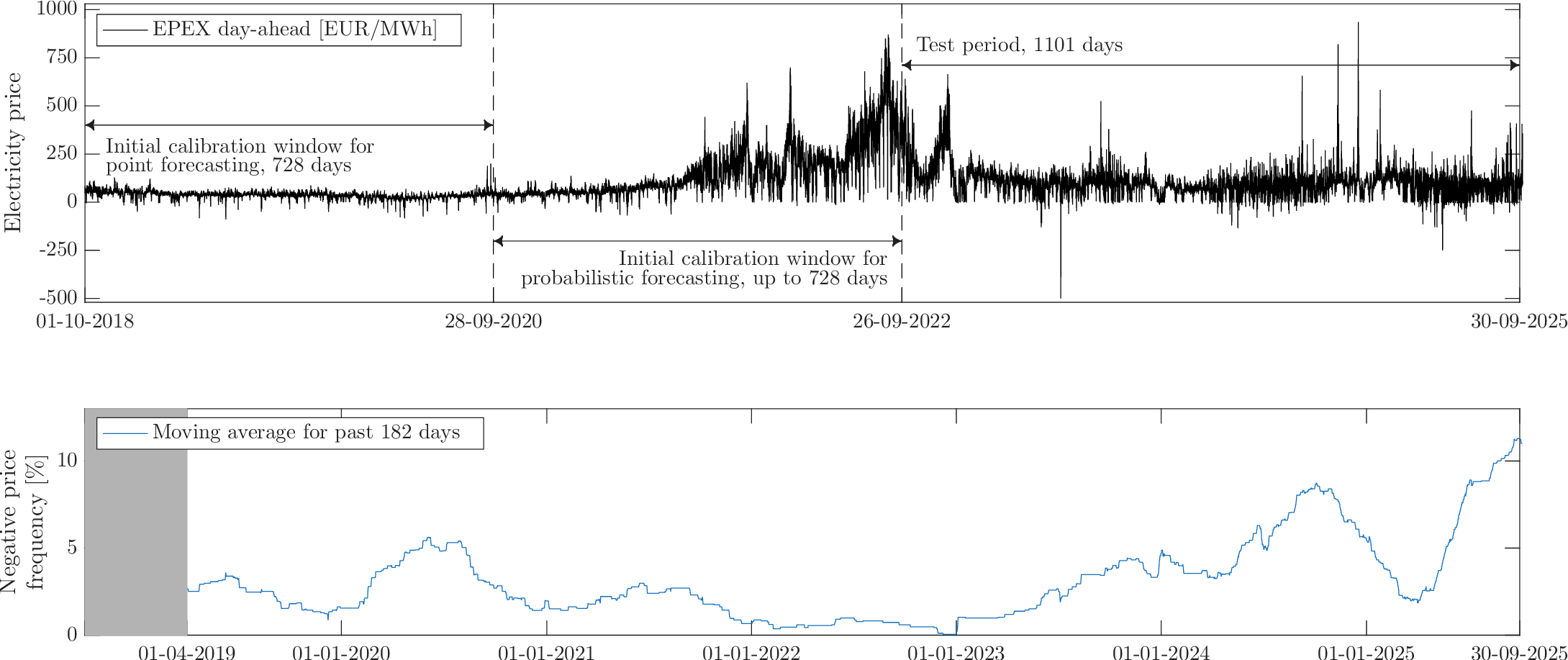}
    \caption{{Hourly German EPEX day-ahead prices (top panel) and the rolling frequency of negative prices computed over a 182-day window (bottom panel). The first and second vertical dashed lines in the top panel indicate the end of the initial calibration period for point and probabilistic forecasting, respectively.}}
    \label{fig:data}
\end{figure*}

\subsection{Test ground}

We analyzed hourly day-ahead electricity prices from the German EPEX SPOT market from 01-10-2018 to 30-09-2025. {The price series was obtained from SMARD \url{https://www.smard.de}}. The German EPEX day-ahead market is a natural empirical setting for this study, as it is one of the standard test beds in the electricity price forecasting literature \citep{cal:fus:ron:17, che:uni:wer:25, lag:der:des:18,m-h:feu:neu:18}. We chose the starting date to align with a structural change in the bidding zone configuration when Austria was excluded from the former DE-AT-LU zone. The end date was set as the day before the day-ahead market transitioned to a 15-minute product.

In addition to price series, we employ day-ahead forecasts of system load and renewable generation. The latter is the sum of wind (onshore and offshore) and solar generation forecasts. {These forecasts} were obtained from the ENTSO-E Transparency Platform \url{https://transparency.entsoe.eu}.
Prior to analysis, all time series were preprocessed to eliminate duplicated or missing observations during daylight saving time transitions.
Furthermore, as the load and renewable forecasts are initially downloaded at a 15-minute resolution, we aggregated them to an hourly frequency to align with the price series.

Figure~\ref{fig:data} displays the hourly price series together with the rolling frequency of negative prices computed over a moving window of 182 days (approximately half a year). The sample contains several episodes of extreme volatility and sustained periods of negative prices. The proportion of negative observations varies over time, with the rolling frequency rising above 10\% at the end of the sample. This time variation provides a suitable environment for assessing performance of post-processing models that must remain coherent when the conditioning forecast becomes negative.

As a robustness check, we also repeat the analysis using data from the
SPPNORTH hub of the Southwest Power Pool (SPP) market for 2019--2025.
Negative prices occur considerably more frequently in SPP than in Germany (see Figure \ref{fig:dataSPP}),
but are substantially less extreme on average. This provides a useful
contrasting setting for assessing the role of both the frequency and the
magnitude of negative-price episodes. The SPP data and results are described
in \ref{sec:spp}.

All forecasting models are estimated using a rolling-window scheme. Point forecast models are re-estimated each day using the most recent 728 days (approximately 2 years) of data. Estimation is performed separately for each delivery hour. This hour-specific, univariate setup follows the standard practice in the electricity price forecasting literature, where forecasts are produced separately for each delivery hour using the same daily information set \citep{han:bar:25}.
Probabilistic postprocessing models are likewise estimated in a rolling manner, using calibration windows of length 182, 364, and 728 days. 
These alternative window lengths allow us to assess the sensitivity of quantile estimation to the amount of recent information incorporated in the calibration sample.

\subsection{Point forecasting models}

To construct inputs for the quantile postprocessing stage, we generate day-ahead point forecasts using three established electricity price forecasting models: ARX, NARX, and LEAR.

Prior to model estimation, prices are first centered using a robust standardization based on the sample median $a$ and median absolute deviation $b$. 
To accommodate heavy tails and negative values, we then apply the inverse hyperbolic sine transformation \citep{sch:11},
\[
y_{d,h} = \operatorname{asinh}\left(\frac{{P}_{d,h}-a}{b}\right),
\]
which has become common in electricity price forecasting due to its variance-stabilizing properties \citep{uni:wer:zie:18}. 

The ARX and NARX specifications share the same set of regressors. 
Autoregressive specifications with exogenous load and renewable information are standard in short-term electricity price forecasting \citep{kel:etal:16,mac:22}.
For each day $d$ and hour $h$, the transformed price is modeled as
\begin{align*}
 y_{d,h} & = \beta_1 y_{d-1,h} + \beta_2 y_{d-2,h} + \beta_3 y_{d-7,h} 
 + \beta_4 y_{d-1,24} + \beta_5 y_{d-1}^{\min}  \nonumber\\ 
 & + \beta_6 y_{d-1}^{\max} + \beta_7 \widehat{\text{Load}}_{d,h} + \beta_8 \widehat{\text{RES}}_{d,h}
 + \sum_{i=1}^7 \beta_{8+i} D_i + \varepsilon_{d,h},
\end{align*}
where $y_d^{\min}$ and $y_d^{\max}$ are, respectively, the lowest and highest values of transformed price on day $d$; $\widehat{\text{Load}}_{d,h}$ and $\widehat{\text{RES}}_{d,h}$ denote day-ahead forecasts of system load and aggregated renewable generation; $D_i$ are weekday indicators; $\varepsilon_{d,h}$ is an error term; and the $\beta_i$ are parameters. 
The ARX model is estimated by ordinary least squares. The NARX model replaces the linear mapping with a feed-forward neural network architecture following the NARX electricity-price forecasting literature \citep{and:atr:09,mar:uni:wer:18}. 
To enhance stability, following \cite{lip:uni:25} NARX forecast is constructed as an ensemble average of ten independently trained networks.

The LEAR model extends this framework by incorporating full daily lag structures of prices, load, and renewable generation \citep{lag:mar:des:wer:21}:
\begin{align*}
 y_{d,h} & = \beta_0 + \sum_{i=1}^{24} \left( \beta_i y_{d-1,i} + \beta_{24+i} y_{d-7,i} \right) 
 + \beta_{49} y_{d-1}^{\min} + \beta_{50} y_{d-1}^{\max} \nonumber\\
 & + \sum_{i=1}^{24} \left( \beta_{50+i} \widehat{\text{Load}}_{d,i} + \beta_{74+i} \widehat{\text{Load}}_{d-1,i} \right) \nonumber\\
 & + \sum_{i=1}^{24} \left(\beta_{98+i} \widehat{\text{RES}}_{d,i} +\beta_{122+i} \widehat{\text{RES}}_{d-1,i}\right) \nonumber\\
 & + \sum_{i=1}^7 \beta_{146+i} D_i + \varepsilon_{d,h}.
\end{align*}

Before entering the postprocessing stage the point forecasts are transformed back to the original price scale. {We also considered postprocessing directly on the transformed scale, but preliminary experiments resulted in substantially too narrow predictive intervals. Moreover, zero on the transformed scale corresponds to the centering parameter rather than to a zero electricity price, making it less suitable for the sign-dependent setting considered here.}

\section{Quantile regression postprocessing}
\label{sec:method}

Our objective is to construct predictive distributions in the form of 99 conditional quantiles corresponding to probability levels 
$\alpha = 0.01, 0.02, \dots, 0.99$. 
Let $\widehat{P}_{d,h}$ denote the day-ahead point forecast for day $d$ and hour $h$. 
Quantile regression postprocessing models the conditional quantile 
$Q_{\alpha}(P_{d,h} \mid \widehat{P}_{d,h})$ 
of the realized price $P_{d,h}$ as a function of the point forecast $\widehat{P}_{d,h}$.

\subsection{Sign-invariant QR postprocessing}
\label{subsec:si}

We begin with the standard single-forecast QR postprocessing specification described in Section~\ref{sec:motivation}. We refer to it as the $sign$-$invariant$ (SI) specification, because the same linear relationship between the point forecast and the conditional quantile is imposed irrespective of whether the point forecast is positive or negative.

In the empirical implementation, the SI specification is estimated separately for each delivery hour. We consider two estimation variants. In the independent variant, denoted SI-I, quantile regressions are estimated separately for each probability level $\alpha$. For a rolling calibration window $\mathcal{W}_d$, the parameters are obtained by the following minimization \citep{koe:05},
\begin{equation}
\min_{a_{\alpha,h},\,b_{\alpha,h}}
\sum_{s\in \mathcal{W}_d}
\rho_{\alpha}
\left(
P_{s,h}
-
a_{\alpha,h}
-
b_{\alpha,h}\widehat{P}_{s,h}
\right),
\qquad
b_{\alpha,h}\geq 0,
\label{eq:si_independent}
\end{equation}
where
\begin{equation*}
    \rho_{\alpha}(u)=u\left(\alpha-\mathbf{1}_{\{u<0\}}\right)
\end{equation*}

is the standard quantile regression pinball loss function {and $W_d$ denotes the rolling calibration window used to estimate the postprocessing model for day $d$.} The non-negativity constraint on $b_{\alpha,h}$ ensures that, for a fixed probability level, higher point forecasts do not imply lower predicted prices \citep{lip:uni:25}.

In the joint variant, denoted SI-J, the quantile regressions for different probability levels are estimated simultaneously for each delivery hour. For computational reasons, joint estimation is performed only for the reduced grid $
\mathcal{A}^{Q}=\{0.01,0.05,0.10,0.15,\ldots,0.95,0.99\}.$
The coefficients for the remaining probability levels in 
$\mathcal{A}=\{0.01,0.02,\ldots,0.99\}$
are obtained by linear interpolation across probability levels. Interpolation is applied to the coefficients rather than to the predicted quantiles.

The objective is the sum of pinball losses over all probability levels in \(\mathcal{A}^{Q}\),
\begin{equation}
\min_{\{a_{\alpha,h},\,b_{\alpha,h}\}_{\alpha\in\mathcal{A}^{Q}}}
\sum_{\alpha\in\mathcal{A}^Q}
\sum_{s\in \mathcal{W}_d}
\rho_{\alpha}
\left(
P_{s,h}
-
a_{\alpha,h}
-
b_{\alpha,h}\widehat{P}_{s,h}
\right),
\label{eq:si_joint}
\end{equation}

subject to the slope non-negativity constraints and the cross-quantile coefficient constraints
\begin{equation}
a_{\alpha_i,h} \leq a_{\alpha_j,h},
\qquad
b_{\alpha_i,h} \leq b_{\alpha_j,h},
\qquad
\text{for all } \alpha_i < \alpha_j .
\label{eq:si_joint_constraints}
\end{equation}

Although SI-J imposes monotonicity on the estimated coefficients, it does not fully eliminate the structural problem discussed in Section~\ref{sec:motivation}. The constraint $b_{\alpha_i,h} \leq b_{\alpha_j,h}$ is natural when the point forecast is positive, but when $\widehat{P}_{d,h}<0$ the slope component works in the opposite direction.

\subsection{Sign-decomposed QR postprocessing}
\label{subsec:sd}

To overcome the problems with the SI specification, we introduce two sign-dependent extensions. The first of these decomposes the point forecast into its positive and negative components,
\[
\widehat{P}^{+}_{d,h}=\max\{0,\widehat{P}_{d,h}\},
\qquad
\widehat{P}^{-}_{d,h}=\min\{0,\widehat{P}_{d,h}\}.
\]
The conditional quantile is then modeled as
\begin{equation}
Q_{\alpha}(P_{d,h}\mid \widehat{P}_{d,h})
=
a_{\alpha,h}
+
b^{+}_{\alpha,h}\widehat{P}^{+}_{d,h}
+
b^{-}_{\alpha,h}\widehat{P}^{-}_{d,h}.
\label{eq:sd}
\end{equation}
We refer to this specification as $sign$-$decomposed$ (SD), because it allows the slope of the quantile function to differ between positive and negative point forecasts, while retaining a common intercept.

In the independent variant of the SD specification, denoted SD-I, the model in Eq.~\eqref{eq:sd} is estimated separately for each probability level $\alpha$ by minimizing the pinball loss. The slope coefficients are constrained to be non-negative,
\[
b^{+}_{\alpha,h}\geq 0,
\qquad
b^{-}_{\alpha,h}\geq 0.
\]
The coefficient $b^{+}_{\alpha,h}$ controls the response of the quantile to positive point forecasts, while $b^{-}_{\alpha,h}$ controls the response to negative point forecasts.

In the joint variant of the SD specification, denoted SD-J, the quantiles on the reduced grid \(\mathcal{A}^{Q}\) are estimated simultaneously for each delivery hour, with the remaining probability levels obtained by coefficient interpolation as in the SI-J specification. The same non-negativity constraints are imposed on the slope coefficients. In addition, the coefficients are constrained across probability levels as follows:
\begin{equation*}
a_{\alpha_i,h}\leq a_{\alpha_j,h},
b^{+}_{\alpha_i,h}\leq b^{+}_{\alpha_j,h},
b^{-}_{\alpha_i,h}\geq b^{-}_{\alpha_j,h},
\quad
\text{for all } \alpha_i<\alpha_j .
\end{equation*}

The opposite direction of the constraint on $b^{-}_{\alpha,h}$ is essential. Since $\widehat{P}^{-}_{d,h}\leq 0$, an increasing sequence of coefficients on the negative component would lower higher quantiles relative to lower quantiles when the point forecast is negative. The reversed ordering prevents this mechanism and makes the coefficient constraints compatible with both positive and negative point forecasts. Thus, unlike SI-J, the SD-J specification can impose monotonicity in a way that accounts for the sign of the conditioning forecast.

\subsection{Sign-specific QR postprocessing}
\label{subsec:ss}

Our second sign-dependent extension allows both the intercept and the slope of the quantile function to differ across positive and negative point forecasts. Instead of decomposing the point forecast into two components within a single regression equation, the sign-specific specification uses separate linear relationships in the two regimes:
\begin{align}
Q_{\alpha}(P_{d,h}\mid \widehat{P}_{d,h})
=
& \mathbf{1}_{\{\widehat{P}_{d,h}\geq 0\}}
\left(
a^{+}_{\alpha,h}
+
b^{+}_{\alpha,h}\widehat{P}^+_{d,h}
\right) \nonumber
\\ 
+&
\mathbf{1}_{\{\widehat{P}_{d,h}<0\}}
\left(
a^{-}_{\alpha}
+
b^{-}_{\alpha}\widehat{P}^-_{d,h}
\right).
\label{eq:ss}
\end{align}
We refer to this specification as $sign$-$specific$ (SS), because the full linear quantile relationship is allowed to differ depending on the sign of the point forecast.

For positive point forecasts, the parameters, $a^{+}_{\alpha,h}$
and $b^{+}_{\alpha,h}$, are estimated separately for each delivery hour, as in the SI and SD specifications. For negative point forecasts, however, observations are pooled across delivery hours within the rolling calibration window, resulting in parameters, $a^{-}_{\alpha}$
and $b^{-}_{\alpha}$, that are common across hours. This is done because negative point forecasts occur relatively infrequently for individual hours, so estimating a separate negative-regime relationship for each hour would often be unstable. {Pooling across hours therefore increases the effective sample size and improves estimation stability, but may introduce bias if the relationship between negative point forecasts and realized prices differs systematically across delivery hours.}\footnote{{We also considered partial pooling across peak and off-peak hours, but the resulting negative-regime subsamples are currently too small to support stable estimation of extreme quantiles. As negative point forecasts become more frequent, block-pooled or hour-specific negative-regime specifications may become feasible and provide a natural extension of the proposed approach.}}

\begin{table}[b!]
\centering
\caption{Summary of QR-based postprocessing specifications.}
\label{tab:qra_models}
\renewcommand{\arraystretch}{1.15}
\scalebox{0.88}{
\begin{tabular}{llll}
\toprule
Model 
& \begin{tabular}{@{}l@{}}Hour\\ treatment\end{tabular} 
& \begin{tabular}{@{}l@{}}Quantile\\ treatment\end{tabular} 
& \begin{tabular}{@{}l@{}}Shape\\ constraints\end{tabular} \\
\midrule

SI-I 
     & separate  & separate 
     & $b_{\alpha,h}\ge0$ \\

\arrayrulecolor{gray!50}
\cmidrule(lr){1-4}
\arrayrulecolor{black}

SI-J 
     & separate  & joint
     & $b_{\alpha,h}\ge0$, $a_{\alpha,h}, b_{\alpha,h} \uparrow$ \\[0.35em]

\midrule
SD-I 
     & separate  & separate 
     & {$b^+_{\alpha,h}, b^-_{\alpha,h}\ge0$} \\

\arrayrulecolor{gray!50}
\cmidrule(lr){1-4}
\arrayrulecolor{black}

SD-J
     & separate  & joint
     & \begin{tabular}{@{}l@{}}
       {$b^+_{\alpha,h}, b^-_{\alpha,h}\ge0$,} \\
       {$a_{\alpha,h}, b^+_{\alpha,h} \uparrow$, 
       $b^-_{\alpha,h} \downarrow$}
       \end{tabular} \\[0.35em]

\midrule

SS-I 
     & \begin{tabular}{@{}l@{}}
       pos.: separate \\ 
       neg.: pooled
       \end{tabular}
     & separate 
     & {$b^+_{\alpha,h}, b^-_{\alpha}\ge0$ }\\

\arrayrulecolor{gray!50}
\cmidrule(lr){1-4}
\arrayrulecolor{black}

SS-J 
     & \begin{tabular}{@{}l@{}}
       pos.: separate \\ 
       neg.: pooled
       \end{tabular}
     & joint
     & \begin{tabular}{@{}l@{}}
       {$b^+_{\alpha,h}, b^-_{\alpha}\ge0$,} \\
       {$a^+_{\alpha,h}, b^+_{\alpha,h} \uparrow$,} \\
       {$a^-_{\alpha} \uparrow$, $b^-_{\alpha} \downarrow$}
       \end{tabular} \\

\bottomrule
\end{tabular}
}
\end{table}

In the independent variant, denoted SS-I, the model in Eq.~\eqref{eq:ss} is estimated separately for each probability level $\alpha$ by minimizing the pinball loss. The slope coefficients are constrained to be non-negative,
\[
b^{+}_{\alpha,h}\geq 0,
\qquad
b^{-}_{\alpha}\geq 0.
\]

In the joint variant, denoted SS-J, the quantiles on the reduced grid \(\mathcal{A}^{Q}\) are estimated simultaneously, with the remaining probability levels obtained by coefficient interpolation. The same slope non-negativity constraints are imposed. For the positive regime, coefficient monotonicity is imposed separately for each delivery hour:
\begin{equation*}
a^{+}_{\alpha_i,h}\leq a^{+}_{\alpha_j,h},
\qquad
b^{+}_{\alpha_i,h}\leq b^{+}_{\alpha_j,h},
\qquad
\text{for all } \alpha_i<\alpha_j .
\label{eq:ss_positive_constraints}
\end{equation*}
For the negative regime, monotonicity is also imposed on the coefficients:
\begin{equation*}
a^{-}_{\alpha_i}\leq a^{-}_{\alpha_j},
\qquad
b^{-}_{\alpha_i}\geq b^{-}_{\alpha_j},
\qquad
\text{for all } \alpha_i<\alpha_j .
\end{equation*}

\begin{figure*}[thb!]
    \centering
    \includegraphics[width=0.75\linewidth]{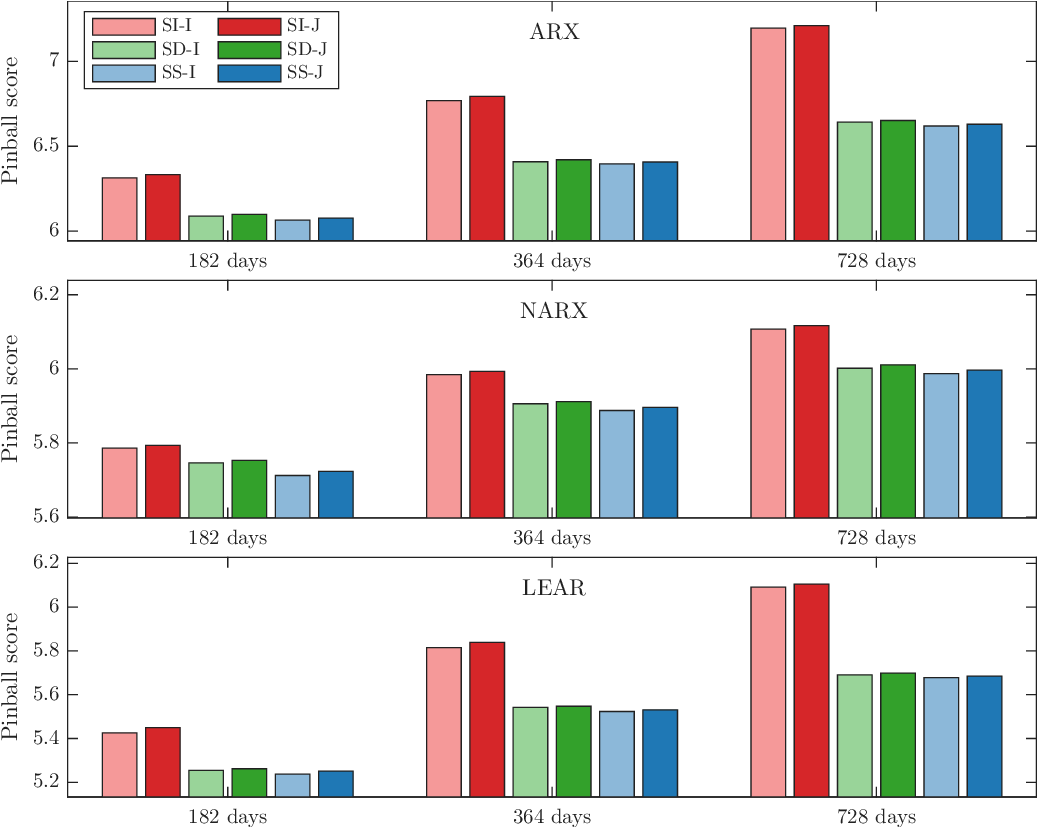}
    \caption{{Average pinball score for the considered QR-based postprocessing specifications for calibration window lengths of 182, 364, and 728 days, and point forecasting models ARX, NARX and LEAR. SI, SD, and SS denote the standard, sign-decomposed, and sign-specific variants, respectively, while -I and -J refer to the estimation schemes. Note: The y-axis scale differs across panels to enhance within-model comparability.}}
    \label{fig:crps}
\end{figure*}

As in the SD-J specification, the ordering of the negative-regime slope coefficients is reversed because the point forecast is negative in this regime. The SS specification is therefore the most flexible of the three approaches considered here: it allows the intercept as well as the slope to differ between positive and negative forecasts. The different specifications are summarized in Table~\ref{tab:qra_models}. 

\section{Empirical results}
\label{sec:results}

This section evaluates the predictive performance of the considered quantile regression postprocessing specifications. {The main results reported in this section refer to the German market. An analogous analysis for the SPP market is reported in \ref{sec:spp} as a robustness exercise.} The analysis proceeds in four steps. First, we compare models in terms of the pinball loss aggregated across all hours and quantiles. Second, we focus on the pinball loss for quantiles out in the tails of the predictive distribution. Third, we restrict attention to observations with negative electricity price forecasts. Fourth, we examine the frequency of quantile crossing as a measure of the internal coherence of the estimated quantile forecasts.

\subsection{Overall predictive accuracy}

Predictive accuracy is assessed using the pinball score \citep[PS;][]{gne:raf:07}, defined for a probability level $\alpha \in (0,1)$ as
\begin{equation}
PS_{\alpha,d,h}
=
\rho_\alpha\left(P_{d,h} - Q_\alpha(P_{d,h}\mid \widehat{P}_{d,h})\right),
\end{equation}
where $P_{d,h}$ denotes the realized price and
$Q_\alpha(P_{d,h}\mid \widehat{P}_{d,h})$ is the corresponding conditional $\alpha$-quantile forecast.

Figure~\ref{fig:crps} reports the average pinball score across all 24 hours and 99 percentiles for the considered QR-based postprocessing specifications applied to the point forecasts from the three models: ARX, NARX and LEAR. The figure reports the results for the three specifications (SI, SD and SS), two estimation schemes (-I and -J), and three calibration window lengths (182, 364 and 728 days).

Several clear patterns emerge from the results in Figure~\ref{fig:crps}. Firstly, the standard specification (SI) produces the poorest predictive results for all three point forecasting models and all three calibration window lengths. This suggests that imposing a single linear relationship between the point forecast and the predictive quantiles is inappropriate when the point forecast can be either positive or negative. Allowing the quantile function to depend on the sign of the point forecast improves predictive accuracy. Figure~\ref{fig:crps} shows that the sign-decomposed specification (SD) reduces the pinball score relative to SI, with the sign-specific specification (SS) providing a further improvement. The gains of the new SD and SS specifications over SI are particularly apparent for ARX, and also for NARX and LEAR for the window lengths of 364 and 728 days.

Secondly, Figure~\ref{fig:crps} shows that for all three specifications (SI, SD and SS) there is very little difference in the results for independent estimation (I) versus joint estimation across quantiles (J). 

Thirdly, the choice of the point forecasting model has a clear impact on the level of the pinball score. LEAR consistently delivers the lowest losses, and ARX the highest losses. Despite these differences in absolute accuracy, the relative ranking of the postprocessing specifications is stable across point forecasting models: SI performs the worst, SD improves upon SI, and SS performs the best.

Finally, the length of the calibration window has a pronounced effect on predictive accuracy. Shorter calibration windows systematically lead to lower pinball scores across all point forecasting models and postprocessing specifications. {This confirms that the choice of calibration-window length can be important for probabilistic electricity price forecasting \citep{ser:uni:wer:19}. In our setting, the superior performance of the 182-day window is consistent with substantial non-stationarity: the frequency and depth of negative prices vary over time (see Figure \ref{fig:data}), so older observations may be less representative of the current price regime.}

In addition to the pinball loss, we briefly assess the empirical coverage of the predictive distributions \citep{gne:kat:14}. Specifically, we consider the 90\% prediction interval defined by the 5\% and 95\% quantiles, and compute the proportion of realizations falling within this interval. The results, reported in \ref{app:sec:cov} (Table \ref{app:tab:cov}), indicate that coverage is broadly similar and close to the ideal of 90\% across model specifications.

\subsection{Tail performance}

\begin{figure*}[thb!]
    \centering
    \includegraphics[width=0.75\linewidth]{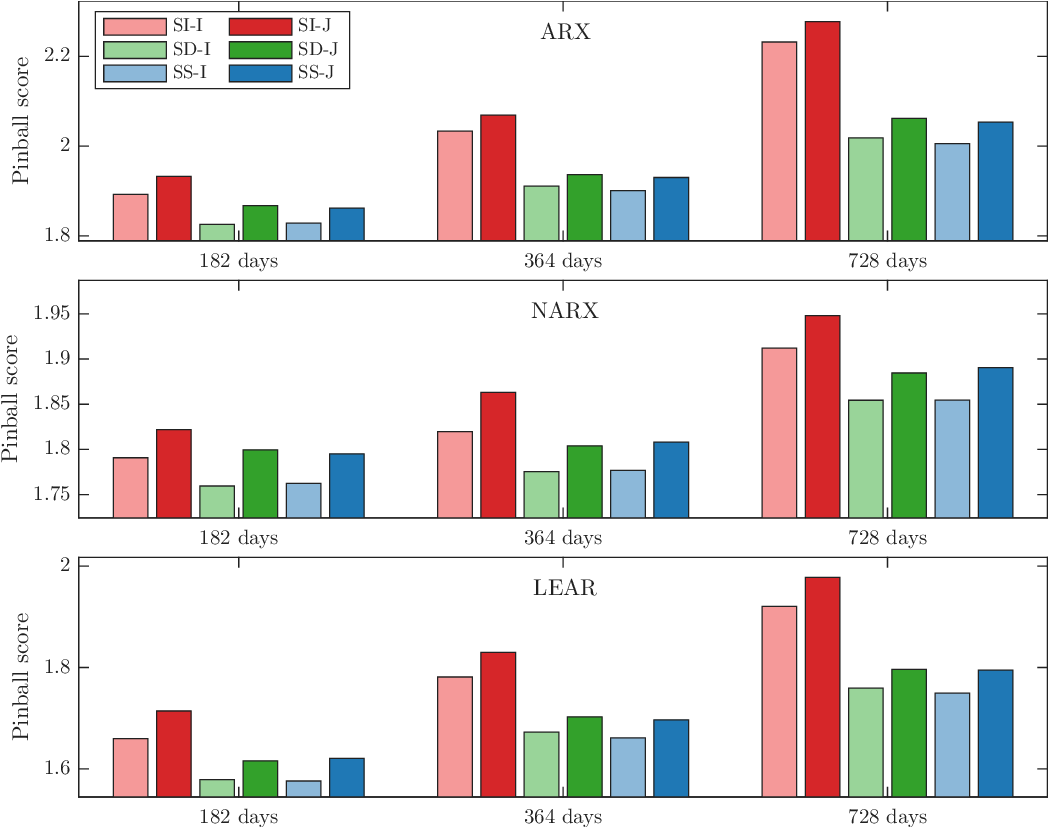}
    \caption{{Average pinball score computed over the 1st--5th and 95th--99th percentile forecasts. The layout and notation follow Figure~\ref{fig:crps}.}}
    \label{fig:crps_tails}
\end{figure*}

We next examine predictive accuracy for the tails of the forecast distribution. The pinball score is averaged over only the extreme probability levels. Specifically, we consider the 1st--5th and 95th--99th percentiles, and average the resulting scores over the 24 hours of the day.

\begin{figure*}[thb!]
    \centering
    \includegraphics[width=0.75\linewidth]{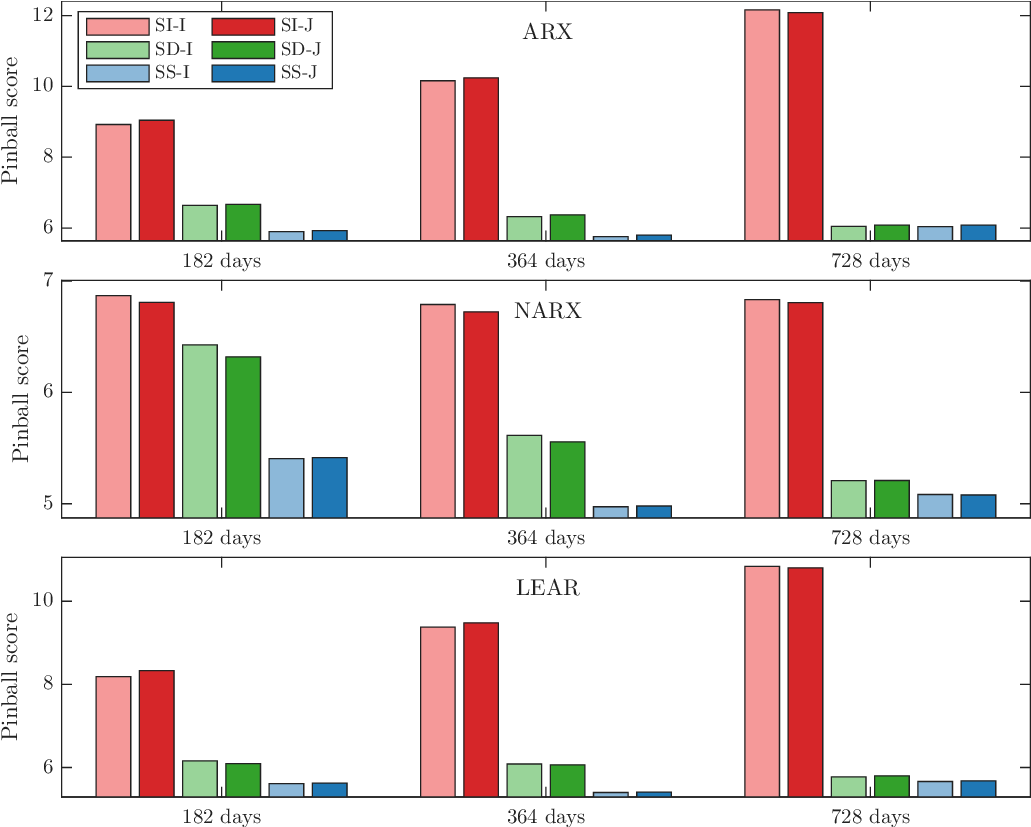}
    \caption{{Average pinball score computed only for observations with negative point forecasts. The layout and notation follow Figure~\ref{fig:crps}.}}
    \label{fig:crps_negatives}
\end{figure*}

Figure~\ref{fig:crps_tails} shows that the results are qualitatively similar to those obtained by averaging across all 99 probability levels. The standard specification (SI) generally produces the least accurate predictions, whereas the sign-dependent specifications improve accuracy. The sign-specific (SS) specification is usually the best-performing approach, although the difference between the sign-dependent (SD) and sign-specific (SS) specifications is very small.

The main difference relative to the full-distribution evaluation is that the gains from sign-dependent postprocessing are more visible for these extreme probability levels, especially for ARX and for longer calibration windows. As before, shorter calibration windows lead to lower pinball scores.

\subsection{Performance for negative price forecasts}

We now restrict the evaluation to observations for which the point forecast is negative.\footnote{The set of observations included in this evaluation depends on the point forecasting model, because each model produces negative forecasts at different times and for different delivery hours. Consequently, the results in this subsection should be compared across postprocessing specifications within a given point forecasting model, rather than across different point forecasting models.} This case is of particular interest because the standard specification imposes a single linear relationship between the point forecast and the predictive quantiles, which can be problematic when the conditioning forecast is negative.

Figure~\ref{fig:crps_negatives} shows that the differences between specifications become much more pronounced when the point forecast is negative. The standard specification (SI) delivers substantially higher pinball scores than the sign-dependent specifications, especially for longer calibration windows. This confirms that the standard postprocessing structure is poorly suited to this part of the forecast space.

Both sign-dependent specifications improve predictive accuracy, but the strongest gains are obtained for the sign-specific specification (SS). In most cases, SS-I and SS-J yield the lowest pinball scores, with only small differences between the two estimation schemes (-I and -J). The sign-decomposed specification (SD) also improves substantially upon SI.

The effect of the calibration window differs from that of the full-sample evaluation. For SI, Figure~\ref{fig:crps_negatives} shows that longer windows tend to worsen performance when forecasts are negative. {This effect is particularly relevant for negative-price forecasts, as both the frequency and the depth of negative-price episodes vary over time. Longer windows may therefore incorporate observations from substantially different negative-price regimes.} For the sign-dependent specifications, and particularly for SS, losses remain much more consistent across different window lengths. Overall, the results for periods with negative point forecasts demonstrate that, for these periods, it is particularly important to explicitly separate the behaviour of positive and negative point forecasts in the postprocessing specification.

\subsection{Internal coherence of predictive distributions}

\begin{figure*}[thb!]
    \centering
    \includegraphics[width=0.75\linewidth]{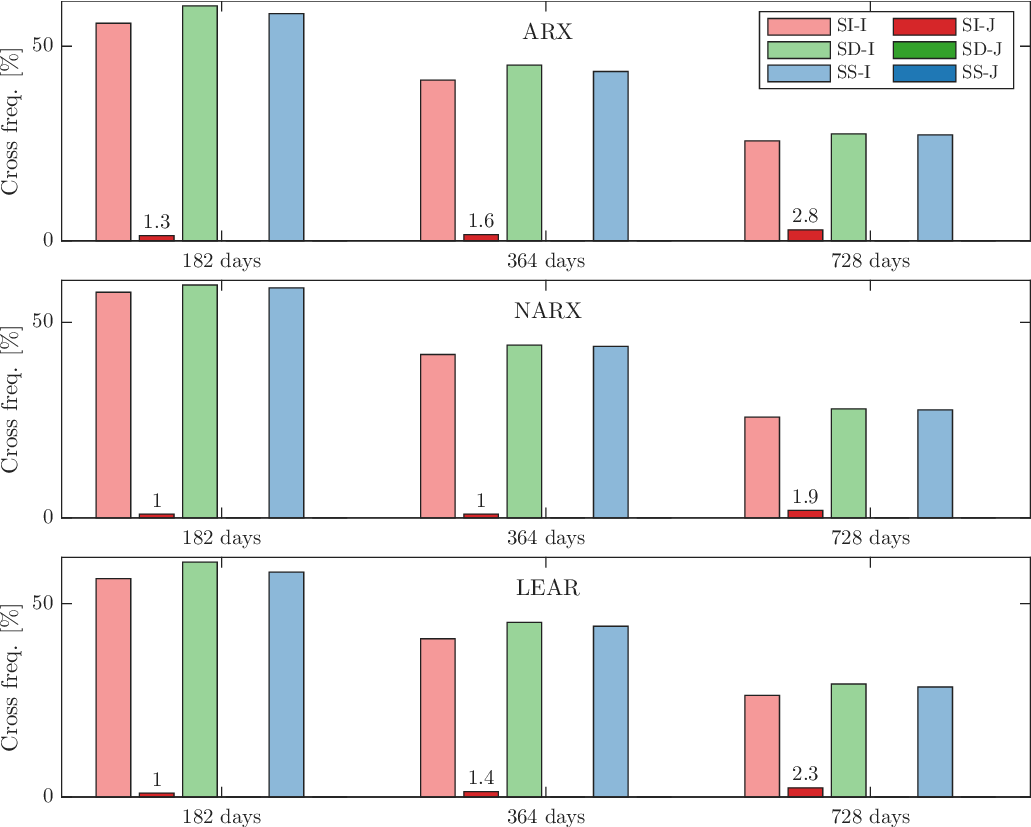}
    \caption{{Crossing frequency for the quantile forecasts. The layout and notation follow Figure~\ref{fig:crps}. Numerical labels are shown for SI-J to highlight that crossing persists under this standard specification even with joint estimation across quantiles.}}
    \label{fig:cross_inc}
\end{figure*}

We now turn from predictive accuracy to the internal coherence of the estimated predictive distributions. A valid set of quantile forecasts should be nondecreasing in the probability level. Therefore, if a lower quantile exceeds a higher quantile, the resulting predictive distribution is internally inconsistent.

It is useful to distinguish between two types of violations. The first is ordinary quantile crossing, which may occur locally between neighboring or nearby probability levels. Such crossings are common in independently estimated quantile regressions and may partly reflect finite-sample or numerical estimation effects. Although undesirable, they can perhaps be corrected by rearranging or sorting the predicted quantiles \citep{che:f-v:gal:10,now:wer:15}. Section 5.4.1 reports results for the frequency of ordinary quantile crossing. 

The second type of quantile crossing is more severe. In the presence of negative point forecasts, the standard specification may produce a reversal of a large part of the predictive distribution, with low quantiles exceeding much higher quantiles. This is not merely a small local crossing problem, but a severe inversion of the predictive distribution induced by the model specification. We refer to such cases as $tail$-$quantile$ $inversion$, and report their frequency in Section 5.4.2.

{The distinction between ordinary quantile crossing and tail-quantile inversion is economically important. Ordinary local crossing may be viewed as a technical defect of the estimated quantile function. Tail-quantile inversion, by contrast, can reverse the interpretation of downside and upside price risk. For example, a storage operator, renewable producer, or electricity trader using the 5\% and 95\% quantiles as risk indicators would receive contradictory signals about the lower and upper tails of the price distribution. Such incoherence can affect bidding strategies, hedging positions, and the valuation of flexibility. Therefore, tail-quantile inversion is not merely a statistical inconvenience, but a market-relevant failure of the probabilistic forecasting system.}

\subsubsection{Ordinary quantile crossing}

We first consider the overall frequency of quantile crossing. This measure records whether at least one crossing occurs in a given hourly predictive distribution, regardless of where in the distribution the crossing takes place.

Let $t=1,\ldots,T$ index all hourly forecasts in the evaluation sample, and let $Q_{\alpha,t}$ denote the raw forecast of the $\alpha$-quantile for observation $t$. We define
\begin{align}
C_t^{\mathrm{any}}
&=
\mathbb{I}\left\{
\exists\, \alpha_i < \alpha_j :
Q_{\alpha_i,t} > Q_{\alpha_j,t}
\right\}, \nonumber \\
\mathrm{Crossing\ frequency}
&=
\frac{1}{T}\sum_{t=1}^T C_t^{\mathrm{any}}.
\end{align}
Thus, the crossing frequency is the share of hourly forecasts for which at least one lower quantile exceeds a higher quantile.

\begin{figure*}[thb!]
    \centering
    \includegraphics[width=0.75\linewidth]{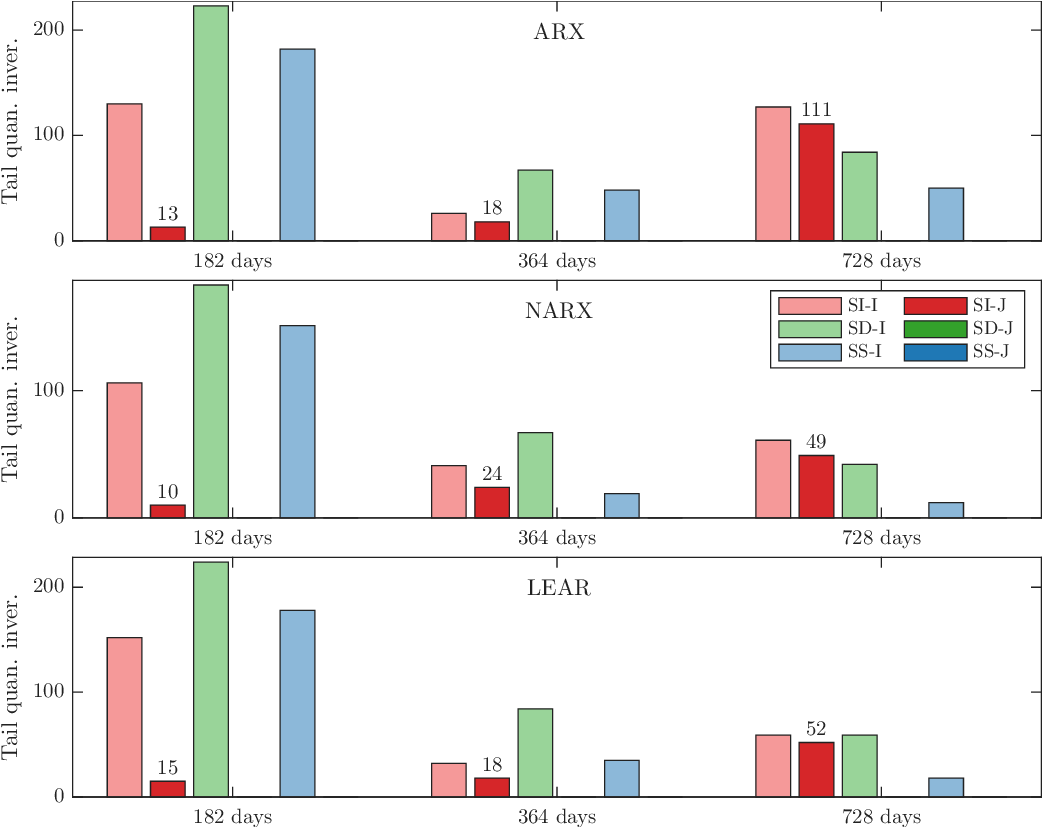}
    \caption{{Tail-quantile inversion count for the raw, unsorted quantile forecasts. The layout and notation follow Figure~\ref{fig:crps}. Numerical labels highlight the non-zero inversion counts for SI-J.}}
    \label{fig:tail_quantile_inversion}
\end{figure*}

Figure~\ref{fig:cross_inc} shows that quantile crossing is very frequent when quantiles are estimated independently (SI-I, SD-I and SS-I). For these specifications, a large share of hourly predictive distributions contain at least one crossing. This pattern is not surprising, as independently estimated quantile regressions do not enforce monotonicity across probability levels. The crossing frequency decreases as the calibration window length increases, suggesting that larger estimation samples stabilize the separately estimated quantile functions.

The more noteworthy outcome relates to the jointly estimated specifications (SI-J, SD-J and SS-J). Since these specifications impose cross-quantile monotonicity restrictions, one would expect quantile crossing to disappear. This is indeed what happens for the sign-dependent specifications: both SD-J and SS-J eliminate quantile crossing in all configurations reported. However, SI-J still produces some crossing, as can be seen from the red bars with numerical labels in Figure~\ref{fig:cross_inc}.

This result illustrates the structural limitation of the standard specification. Under SI, the same linear relationship between the point forecast and the predictive quantiles is imposed regardless of the sign of the forecast. {As shown analytically with reference to Eq.~\eqref{eq:quantile_diff}}, when the point forecast is negative, the ordering induced by the slope coefficients may be reversed, so coefficient monotonicity across probability levels is no longer sufficient to guarantee monotonicity of the resulting quantile forecasts. Hence, the remaining crossings under SI-J are not merely a numerical artifact. They indicate that joint estimation alone cannot fully resolve the coherence problem unless the specification also accounts for the sign of the point forecast.

\subsubsection{Tail-quantile inversion}

{We next consider a more severe form of quantile crossing, which we refer to as tail-quantile inversion. This measure records whether the lower part of the predictive distribution is placed above the upper part. Specifically, we check whether any quantile from the lower tail, defined as probability levels from 1\% to 25\%, exceeds any quantile from the upper tail, defined as probability levels from 75\% to 99\%. Formally,
\begin{equation}
C_t^{\mathrm{tail}}
=
\mathbb{I}\left\{
\exists\, \alpha_i \in \mathcal{A}^{L} \quad \exists\, \alpha_j \in \mathcal{A}^{U}:
Q_{\alpha_i,t} > Q_{\alpha_j,t}
\right\},
\end{equation}
where $\mathcal{A}^{L}=\{0.01,\ldots,0.25\}$ and $\mathcal{A}^{U}=\{0.75,\ldots,0.99\}$.
}
We then report the total number of such cases:
\begin{equation}
\mathrm{Tail\text{-}quantile\ inversion\ count}
=
\sum_{t=1}^T C_t^{\mathrm{tail}}.
\end{equation}

Figure~\ref{fig:tail_quantile_inversion} shows that tail-quantile inversion occurs mainly when quantiles are estimated independently (SI-I, SD-I and SS-I). These variants can produce a substantial number of such cases, especially for the shortest calibration window. This confirms that independent estimation may lead not only to local crossings, but also to severe reversals of the predictive distribution.

The most important result again concerns the jointly estimated specifications (SI-J, SD-J and SS-J). Tail-quantile inversion is completely eliminated for SD-J and SS-J across all point forecasting models and calibration windows. In contrast, SI-J still produces some inversions, as highlighted by the numerical labels on the red bars in Figure~\ref{fig:tail_quantile_inversion}. Although these counts are relatively small their presence is important: under joint estimation, such inversions would be expected to disappear if the specification itself were coherent.

\section{Economic evaluation}
\label{sec:economic}

\begin{figure*}[thb!]
    \centering
    \includegraphics[width=0.75\textwidth]{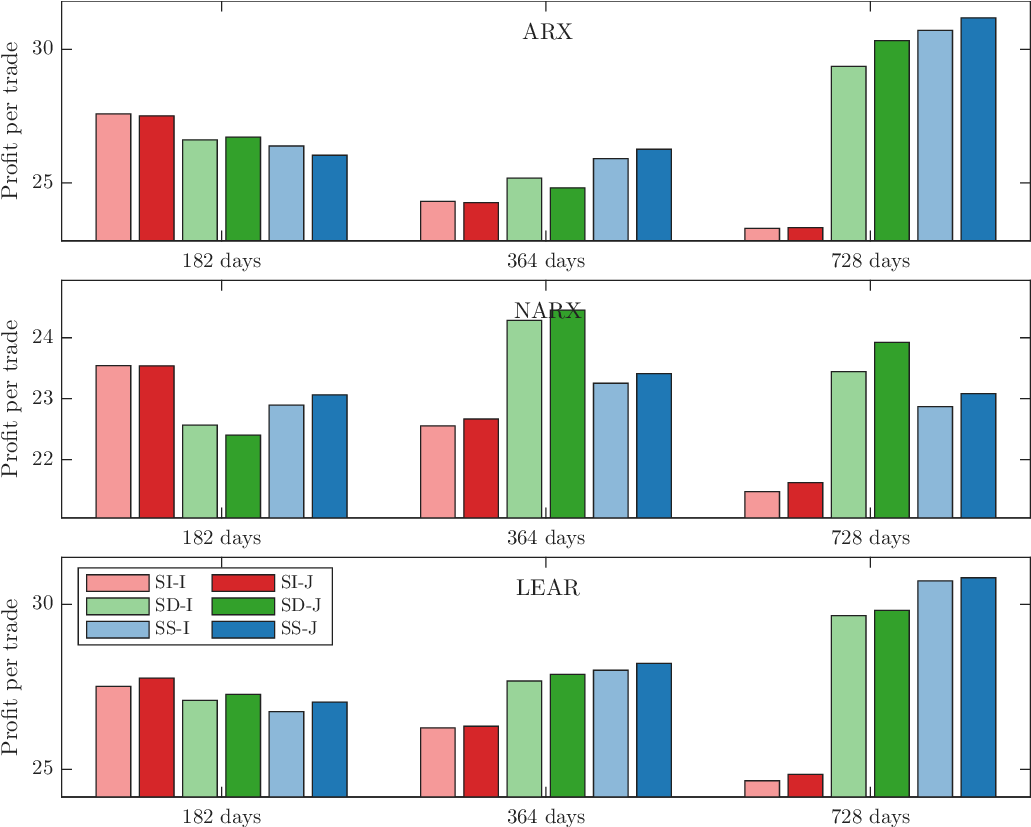}
    \caption{{Average profit per executed trade for the battery arbitrage strategy with risk threshold $\pi=1/3$. For comparison, the benchmark strategy based solely on point forecasts, without using probabilistic information, yields average profits per trade of 20.72, 19.32, and 19.16 EUR/MWh for ARX, NARX, and LEAR, respectively. The layout and notation follow Figure~\ref{fig:crps}.}}
    \label{fig:profit}
\end{figure*}

{Statistical scores provide information about the quality of probabilistic forecasts, but do not directly indicate whether improvements in forecast
accuracy translate into better economic decisions. In electricity markets, the ranking of probabilistic forecasts according to statistical criteria need not coincide with their usefulness in decision-oriented applications
\citep{nit:wer:23}. We therefore complement the statistical evaluation with a simple day-ahead battery arbitrage strategy following
\cite{lip:uni:25}.

We consider a battery energy storage system that performs at most one
charge--discharge cycle per day. Before applying the trading strategy, the 99 predicted quantiles are sorted in ascending order for each day and
delivery hour. This ensures that all trading decisions are based on monotone predictive distributions, including those obtained from specifications that
produce quantile crossing.

For each day $d$, the buying hour $h_b$ and the subsequent selling hour
$h_s$ are selected using the median forecasts. More precisely, the pair
$(h_b,h_s)$ is chosen to maximize the predicted spread, and the corresponding
net gain is
\begin{equation}
\widehat{S}_d =
\eta Q_{0.5}(P_{d,h_s}\mid\widehat{P}_{d,h_s})
-
\frac{1}{\eta}
Q_{0.5}(P_{d,h_b}\mid\widehat{P}_{d,h_b})
-C,
\label{eq:predicted_profit}
\end{equation}
where $h_b<h_s$, $\eta$ denotes the battery efficiency and $C$ is the
operating cost of one complete charge--discharge cycle. Following \cite{ser:wer:25} we set
$\eta=0.9$ and $C=100$ EUR. The latter
provides an approximation of the degradation and operating costs associated
with a utility-scale lithium-ion battery.

The median forecast determines the trading opportunity, while the remaining quantiles provide information about the associated uncertainty. At the buying hour, the unfavourable outcome is that the price is higher than the median forecast suggests. Conversely, at the selling hour, the unfavourable outcome is a lower price than the median forecast. Following \cite{lip:uni:25}, we quantify these risks using the mean upper and lower partial deviations from the corresponding median forecasts.

\begin{equation*}
\begin{split}
R_d &=
\frac{\eta}{49}
\sum_{i=1}^{49}
\left[
Q_{0.5}(P_{d,h_s}\mid\widehat{P}_{d,h_s})
-
Q_{\frac{i}{100}}(P_{d,h_s}\mid\widehat{P}_{d,h_s})
\right]
\\
&+
\frac{1}{49\eta}
\sum_{i=1}^{49}
\left[
Q_{1-\frac{i}{100}}(P_{d,h_b}\mid\widehat{P}_{d,h_b})
-
Q_{0.5}(P_{d,h_b}\mid\widehat{P}_{d,h_b})
\right].
\end{split}
\label{eq:trading_risk}
\end{equation*}
Thus, $R_d\in\mathbb{R}_{+}$ summarizes the downside risk at the selling
hour and the upside risk at the buying hour.

A trade is executed only when the predicted net gain is sufficiently large
relative to its associated uncertainty,
\begin{equation}
\widehat{S}_d > \pi R_d.
\label{eq:trade_rule}
\end{equation}
We set the risk threshold to $\pi=1/3$. The purpose of this exercise is to
compare the economic usefulness of the considered probabilistic forecasts,
rather than to optimize the trading rule, and we therefore use the same
fixed threshold for all models and postprocessing specifications.

If the condition in Eq.~(\ref{eq:trade_rule}) is satisfied, the realized
profit is calculated using the observed prices,
\begin{equation}
S_d =
\eta P_{d,h_s}
-
\frac{1}{\eta}P_{d,h_b}
-C.
\label{eq:realized_profit}
\end{equation}
Otherwise, no transaction is made. Since the different postprocessing
specifications can result in different numbers of executed trades, we
compare their economic performance using the average profit per trade (EUR/MWh)
rather than cumulative profit.

For reference, we also perform the test for a benchmark strategy based solely on the underlying point forecasts, with a trade executed whenever the predicted spread is positive. The benchmark strategy yields average profits per trade of 20.72, 19.32, and 19.16 EUR/MWh for ARX, NARX, and LEAR, respectively. All probabilistic strategies reported in Figure~\ref{fig:profit} improve upon these benchmarks, highlighting the value of incorporating forecast uncertainty into the trading decision.

Figure~\ref{fig:profit} shows that the sign-dependent specifications can
translate into economically meaningful gains. Overall, the advantage of SD
and SS over the standard SI specification becomes stronger as the calibration
window increases and is most pronounced for the 728-day window.

Looking first at the postprocessing specifications, SD and SS clearly
outperform SI for all three point forecasting models with the 728-day window.
The improvement is particularly large for ARX and LEAR. With the 364-day
window, the differences are smaller, but the sign-dependent specifications
{consistently outperform SI across all three point forecasting models}.
For the 182-day window, this pattern is not present: SI performs best for {all models but the differences are relatively
small}. Another observation is that the independent and joint estimation variants generally produce similar profits, with a small
advantage of {joint} estimation.

The calibration-window length has a stronger effect on profitability.
Interestingly, the highest profits are generally obtained with the 728-day
window for SD and SS, {except for NARX, where both specifications
perform best with the 364-day window}, despite the shorter windows
performing better in terms of the pinball score.
The ranking of the point forecasting models also differs from that based on
statistical accuracy: ARX, which performs worst in terms of the pinball score,
{outperforms NARX in terms of profits in all configurations and even
yields the highest average profit per trade in a few cases, while LEAR
generally delivers the highest profits and performs best statistically}.
This further illustrates that statistical accuracy and economic value
need not coincide. However, since the models may generate different
numbers of trades, this should not be interpreted as {any single
model being uniformly superior}.

\section{Conclusion}
\label{sec:conclusion}

This paper studies QR-based probabilistic postprocessing in the presence of negative electricity price point forecasts. We show that the standard specification, which imposes a single linear relationship between the point forecast and all predictive quantiles, can lead to internally incoherent predictive distributions when the point forecast takes negative values. In such cases, the problem is not limited to quantile crossing for probability levels that are quite close, but may involve a substantial reversal of the predictive distribution. {From an energy-market perspective, such reversals are problematic because lower and upper quantiles are often interpreted as downside and upside price-risk indicators.}

To address this issue, we consider sign-dependent postprocessing specifications that allow the relationship between the point forecast and the predictive quantiles to differ depending on the sign of the point forecast. The empirical results show that this modification improves predictive accuracy, both for the full set of quantiles and for the tails of the predictive distribution. The gains are particularly pronounced when the evaluation is restricted to observations with negative point forecasts, where the standard specification performs poorly. {This suggests that negative-price regimes should be treated explicitly when calibrating predictive distributions for electricity-market applications.}

The analysis of internal coherence provides an additional argument in favor of the proposed specifications. Independent estimation of quantiles leads to frequent crossing, which is a common issue in separately estimated quantile regressions. More importantly, however, joint estimation alone is not sufficient to ensure coherent predictive distributions under the standard specification. Even with cross-quantile monotonicity constraints, joint estimation of the standard specification (SI-J) continues to produce non-zero crossing frequencies and tail-quantile inversions. In contrast, the sign-dependent specifications combined with joint estimation (SD-J and SS-J) eliminate both ordinary quantile crossing and tail-quantile inversion across all considered point forecasting models and calibration windows. {Thus, the empirical evidence shows that coherence requires not only monotonicity constraints, but also a specification that respects the sign-dependent nature of the price-forecast relationship.}

{The robustness analysis for the SPP market leads to similar conclusions
regarding forecast coherence. The improvements in predictive accuracy are less
pronounced than for Germany, but SD-J and SS-J again eliminate the crossing and
tail-quantile inversion observed under the standard specification. The
comparison between the two markets also indicates that the relevance of the
problem depends on both the frequency and the magnitude of negative prices:
SPP exhibits negative prices more frequently, whereas negative prices in
Germany are considerably more extreme.}

These findings have practical implications for probabilistic electricity price forecasting. Incoherent predictive distributions may provide misleading signals in applications such as trading, bidding, and risk management. {They may also distort the valuation of flexibility, since storage operators, demand-response providers, and renewable producers rely on distributional forecasts to assess downside exposure and upside opportunities.} {This practical relevance is also reflected in the economic evaluation: using probabilistic forecasts in the battery arbitrage strategy increases average profit per trade relative to the point-forecast benchmark, while the sign-dependent specifications can yield further gains.}

{Taken together, our results indicate that explicitly accounting for the sign of the point forecast can improve predictive accuracy, preserve the internal coherence of the resulting predictive distributions, and enhance their economic usefulness. The proposed sign-dependent specifications extend directly to quarter-hourly forecasting. The larger number of delivery periods increases the dimensionality of the forecasting problem, while pooling negative-regime observations across delivery periods remains a natural way to maintain estimation stability.} {As negative prices become more relevant in renewable-based power systems, sign-dependent postprocessing offers a simple way to adapt existing point-forecasting systems to market conditions in which conventional positive-price relationships no longer apply.}

\section*{Acknowledgments}
The study was partially supported by the National Science Center (NCN, Poland) through grant no.\ 2023/49/N/HS4/02741 (to BU).

The majority of this work was developed during BU’s research stay at the Sa\"{\i}d Business School, University of Oxford, United Kingdom, where the main idea of the study was conceived and the core results were obtained.

\section*{Declaration of generative AI}
During the preparation of this work, the authors used ChatGPT (OpenAI) for limited language editing. After using this tool, the authors reviewed and edited the content as needed and take full responsibility for the content of the published article.

\bibliographystyle{elsarticle-harv}
\bibliography{references}

@ARTICLE{and:atr:09,
	author = {Arash Andalib and Farid Atry},
	year = 2009,
	title = {Multi-step ahead forecasts for electricity prices using {NARX}: {A} new approach, a critical analysis of one-step ahead forecasts}, 
	journal = {Energy Conversion and Management}, 
	volume = 50, 
	pages = {739-747}
}

@ARTICLE{cal:fus:ron:17,
	author={Caldana, R. and Fusai, G. and Roncoroni, A.},
	title={Electricity forward curves with thin granularity: Theory and empirical evidence in the hourly EPEXspot market},
	journal={European Journal of Operational Research},
	year={2017},
	volume={261},
	number={2},
	pages={715-734}
}

@ARTICLE{che:f-v:gal:10,
	author={Chernozhukov, V. and Fernandez-Val, I. and Galichon, A.},
	title={Quantile and probability curves without crossing},
	journal={Econometrica},
	year={2010},
	volume={78},
	number={3},
	pages={1093-1125},
}

@ARTICLE{gne:kat:14,
	author={Gneiting, T. and Katzfuss, M.},
	title={Probabilistic forecasting},
	journal={Annual Review of Statistics and Its Application},
	year={2014},
	volume={1},
	pages={125-151}
}

@ARTICLE{gne:raf:07,
	author = {T. Gneiting and A. Raftery},
	year = 2007,
	title = {Strictly Proper Scoring Rules, Prediction,	and Estimation},
	journal = {Journal of the American Statistical Association},
	volume = 102,
	number = 477,
	pages = {359-378}
}

@article{hui:ste:22,
title = {The dependence of quantile power prices on supply from renewables},
journal = {Energy Economics},
volume = {105},
pages = {105685},
year = {2022},
author = {Ronald Huisman and Cristian Stet},
}

@article{jan:woj:22,
title = {Dynamic short-term risk management strategies for the choice of electricity market based on probabilistic forecasts of profit and risk measures. {T}he {G}erman and the {P}olish market case study},
journal = {Energy Economics},
volume = {110},
pages = {106015},
year = {2022},
author = {Joanna Janczura and Edyta Wójcik},
}

@ARTICLE{kel:etal:16,
	author = {Dogan Keles and Jonathan Scelle and Florentina Paraschiv and Wolf Fichtner},
	year = 2016,
	title = {Extended forecast methods for day-ahead electricity spot prices applying artificial neural networks},
	journal = {Applied Energy},
	volume = 162,
	pages = {218-230}
}

@BOOK{koe:05,
	author = {R. W. Koenker},
	year = 2005,
	title = {Quantile Regression},
	publisher = {Cambridge University Press}
}

@ARTICLE{lag:der:des:18,
	author={Lago, J. and De Ridder, F. and De Schutter, B.},
	title={Forecasting spot electricity prices: Deep learning approaches and empirical comparison of traditional algorithms},
	journal={Applied Energy},
	year={2018},
	volume={221},
	pages={386-405}
}

@ARTICLE{lag:mar:des:wer:21,
    author={Lago, J. and Marcjasz, G. and De Schutter, B. and Weron, R.},
    title={Forecasting day-ahead electricity prices: A review of state-of-the-art algorithms, best practices and an open-access benchmark},
    journal={Applied Energy},
    year={2021},
    volume={293},
    pages={116983}
}

@article{lip:uni:wer:24,
title = {Postprocessing of point predictions for probabilistic forecasting of day-ahead electricity prices: The benefits of using isotonic distributional regression},
journal = {Energy Economics},
volume = {139},
pages = {107934},
year = {2024},
author = {Arkadiusz Lipiecki and Bartosz Uniejewski and Rafał Weron},
}

@ARTICLE{mac:20,
	author={Maciejowska, K.},
	title={Assessing the impact of renewable energy sources on the electricity price level and variability -- A quantile regression approach},
	journal={Energy Economics},
	year={2020},
	volume={85},
	pages={104532}
}

@ARTICLE{mac:22,
	author = {K. Maciejowska},
	year = 2022,
	title = {Portfolio management of a small RES utility with a structural vector autoregressive model of electricity markets in {G}ermany},
	journal = {Operations Research and Decisions},
	volume = 32,
	number = 4,
	pages = {75-90}
}

@ARTICLE{mar:uni:wer:18,
	author = {G. Marcjasz and B. Uniejewski and R. Weron},
	year = 2019,
	title = {On the importance of the long-term seasonal component in day-ahead electricity price forecasting with {NARX} neural networks},
	journal = {International Journal of Forecasting},
	volume = 35,
	pages = {1520-1532}
}

@ARTICLE{m-h:feu:neu:18,
	author={M\"arkle-Hu{\ss}, J. and Feuerriegel, S. and Neumann, D.},
	title={Contract durations in the electricity market: Causal impact of 15 min trading on the {EPEX SPOT} market},
	journal={Energy Economics},
	year={2018},
	volume={69},
	pages={367-378}
}

@article{nar:zie:22,
  title={Optimal bidding in hourly and quarter-hourly electricity price auctions: Trading large volumes of power with market impact and transaction costs},
  author={Narajewski, Micha{\l} and Ziel, Florian},
  journal={Energy Economics},
  volume={110},
  pages={105974},
  year={2022},
}

@ARTICLE{nit:wer:23,
	author={Nitka, W. and Weron, R.},
	title={Combining predictive distributions of electricity prices. Does minimizing the {CRPS} lead to optimal decisions in day-ahead bidding?},
	journal={Operations Research and Decisions},
	year={2023},
	volume={33},
    issue=3,
	pages={105-118}
}

@ARTICLE{now:wer:15,  
	author = {J. Nowotarski and R. Weron},
	year = 2015,
	title = {Computing electricity spot price prediction intervals using quantile regression and forecast averaging},
	journal = {Computational Statistics},
	volume = 30,
	number = 3,
	pages = {791-803}
}

@ARTICLE{now:wer:18,
	author = {J. Nowotarski and R. Weron}, 
	year = 2018,
	title = {Recent advances in electricity price forecasting: {A} review of
	probabilistic forecasting},
	journal = {Renewable and Sustainable Energy Reviews},
	volume = 81,
	issue = 1,
	pages = {1548-1568}
}

@ARTICLE{ser:uni:wer:19,
	author={Serafin, T. and Uniejewski, B. and Weron, R.},
	title={Averaging predictive distributions across calibration windows for day-ahead electricity price forecasting},
	journal={Energies},
	year={2019},
	volume={12},
	number={13},
	pages={2561}
}

@ARTICLE{sch:11,
	author={S. Schneider},
	title={Power spot price models with negative prices},
	journal={Journal of Energy Markets},
	year={2011},
	volume={4},
	number={4},
	pages={77-102},
}

@article{uni:wer:zie:18,
	author = {B. Uniejewski and R. Weron and F. Ziel},
	year = 2018,
	title = {Variance Stabilizing Transformations for Electricity Spot Price Forecasting},
	journal={IEEE Transactions on Power Systems},
	volume = 33,
	issue = 2,
	pages = {2219-2229}
}

@article{che:uni:wer:25,
title = {Extrapolating the long-term seasonal component of electricity prices for forecasting in the day-ahead market},
journal = {Journal of Commodity Markets},
volume = {37},
pages = {100449},
year = {2025},
author = {Katarzyna Ch\k{e}ć and Bartosz Uniejewski and Rafał Weron},
}

@misc{lip:uni:25,
      title={Isotonic Quantile Regression Averaging for uncertainty quantification of electricity price forecasts}, 
      author={Arkadiusz Lipiecki and Bartosz Uniejewski},
      year={2025},
      archivePrefix={arXiv},
      url={https://arxiv.org/abs/2507.15079}, 
}

@ARTICLE{fan:gam:pro:13,
  author = {Fanone, Enzo and Gamba, Andrea and Prokopczuk, Marcel},
  title = {The Case of Negative Day-Ahead Electricity Prices},
  journal = {Energy Economics},
  year = {2013},
  volume = {35},
  pages = {22--34},
}

@ARTICLE{wal:hen:zie:gne:24,
  author = {Walz, Eva-Maria and Henzi, Alexander and Ziegel, Johanna F. and Gneiting, Tilmann},
  title = {{Easy Uncertainty Quantification (EasyUQ)}: Generating Predictive Distributions from Single-Valued Model Output},
  journal = {SIAM Review},
  year = {2024},
  volume = {66},
  number = {1},
  pages = {91--122},
}

@INPROCEEDINGS{rom:pat:can:19,
  author = {Romano, Yaniv and Patterson, Evan and Cand{\`e}s, Emmanuel J.},
  title = {Conformalized Quantile Regression},
  booktitle = {Advances in Neural Information Processing Systems},
  year = {2019},
  volume = {32}
}

@ARTICLE{hen:zie:gne:21,
  author = {Henzi, Alexander and Ziegel, Johanna F. and Gneiting, Tilmann},
  title = {Isotonic Distributional Regression},
  journal = {Journal of the Royal Statistical Society: Series B (Statistical Methodology)},
  year = {2021},
  volume = {83},
  number = {5},
  pages = {963--993},
}

@ARTICLE{bon:rei:wan:10,
  author = {Bondell, Howard D. and Reich, Brian J. and Wang, Huixia},
  title = {Noncrossing Quantile Regression Curve Estimation},
  journal = {Biometrika},
  year = {2010},
  volume = {97},
  number = {4},
  pages = {825--838},
}

@ARTICLE{det:vol:08,
  author = {Dette, Holger and Volgushev, Stanislav},
  title = {Non-Crossing Non-Parametric Estimates of Quantile Curves},
  journal = {Journal of the Royal Statistical Society: Series B (Statistical Methodology)},
  year = {2008},
  volume = {70},
  number = {3},
  pages = {609--627},
}

@ARTICLE{can:18,
  author = {Cannon, Alex J.},
  title = {Non-crossing Nonlinear Regression Quantiles by Monotone Composite Quantile Regression Neural Network, with Application to Rainfall Extremes},
  journal = {Stochastic Environmental Research and Risk Assessment},
  year = {2018},
  volume = {32},
  number = {11},
  pages = {3207--3225},
}

@INPROCEEDINGS{par:mad:aub:kan:gas:wan:22,
  author = {Park, Youngsuk and Maddix, Danielle C. and Aubet, Fran{\c c}ois-Xavier and Kan, Kelvin and Gasthaus, Jan and Wang, Yuyang},
  title = {Learning Quantile Functions without Quantile Crossing for Distribution-Free Time Series Forecasting},
  booktitle = {Proceedings of the 25th International Conference on Artificial Intelligence and Statistics},
  series = {Proceedings of Machine Learning Research},
  volume = {151},
  pages = {8127--8150},
  year = {2022}
}

@ARTICLE{nic:10,
  author = {Nicolosi, Marco},
  title = {Wind Power Integration and Power System Flexibility: An Empirical Analysis of Extreme Events in Germany under the New Negative Price Regime},
  journal = {Energy Policy},
  year = {2010},
  volume = {38},
  number = {11},
  pages = {7257--7268},
}

@article{koe:hal:01,
  author  = {Koenker, Roger and Hallock, Kevin F.},
  title   = {Quantile Regression},
  journal = {Journal of Economic Perspectives},
  year    = {2001},
  volume  = {15},
  number  = {4},
  pages   = {143--156}
}

@article{han:bar:25,
title = {Learning the probability distributions of day-ahead electricity prices},
journal = {Energy Economics},
volume = {152},
pages = {108988},
year = {2025},
author = {Luboš Hanus and Jozef Baruník},
}

@article{ser:wer:25,
    author  = {Serafin, Tomasz and Weron, Rafał},
    title   = {Loss functions in regression models: Impact on profits and risk in day-ahead electricity trading},
    journal = {Energy Economics},
    volume  = {148},
    pages   = {108596},
    year    = {2025},
}

@article{leb:das:pap:sch:26,
  author  = {Lebedev, Andreas and Das, Abhinav and Pappert, Sven and Schl{\"u}ter, Stephan},
  title   = {Analyzing Uncertainty Quantification in Statistical and Deep Learning Models for Probabilistic Electricity Price Forecasting},
  journal = {IEEE Access},
  volume  = {14},
  pages   = {52162--52189},
  year    = {2026},
}

\appendix
\setcounter{table}{0}

\section{Empirical Coverage}
\label{app:sec:cov}

\begin{table*}[thb!]
\caption{{Empirical coverage of the 90\% prediction interval for the considered QR-based postprocessing specifications, computed for all observations. Coverage is defined as the percentage of realizations falling within the interval between the 5th and 95th predicted quantiles.}}
\label{app:tab:cov}
\centering
\begin{tabular}{r|ccc|ccc|ccc}
     & \multicolumn{3}{c|}{182 days}  & \multicolumn{3}{c|}{364 days}  & \multicolumn{3}{c}{728 days}  \\
     & ARX   & NARX  & LEAR  & ARX   & NARX  & LEAR  & ARX   & NARX  & LEAR  \\
     \midrule
SI-I & 88.0\% & 88.7\% & 89.1\% & 90.0\% & 90.6\% & 90.8\% & 91.4\% & 91.3\% & 92.4\% \\
  SI-J & 88.1\% & 88.8\% & 89.2\% & 89.9\% & 90.6\% & 90.9\% & 91.4\% & 91.3\% & 92.2\% \\
  SD-I & 88.2\% & 88.7\% & 89.1\% & 90.1\% & 90.4\% & 91.0\% & 91.9\% & 91.4\% & 92.8\% \\
  SD-J & 88.2\% & 88.8\% & 89.3\% & 90.1\% & 90.5\% & 91.2\% & 92.1\% & 91.4\% & 92.8\% \\
  SS-I & 88.1\% & 88.8\% & 89.0\% & 90.2\% & 90.5\% & 91.2\% & 92.1\% & 91.5\% & 92.8\% \\
  SS-J & 88.5\% & 88.9\% & 89.4\% & 90.2\% & 90.7\% & 91.1\% & 92.1\% & 91.5\% & 92.8\% \\
\end{tabular}
\end{table*}

{Table~\ref{app:tab:cov} reports the empirical coverage of the 90\% prediction interval defined by the 5\% and 95\% quantiles. Coverage is very similar across the SI, SD, and SS specifications and between independent and joint estimation, indicating that the proposed modifications do not strongly affect the overall calibration of the prediction intervals. The main difference is associated with the calibration-window length: the 182-day window yields coverage of around 88--89\%, indicating slightly too narrow intervals, whereas the longer windows provide higher coverage that is generally close to the nominal 90\% level.}

{
\section{Robustness checks}
\label{sec:spp}

\subsection{SPP data}
\label{app:spp_data}

The robustness analysis uses hourly day-ahead prices from the Southwest Power
Pool (SPP) for 2019--2025. Prices are represented by the Locational Marginal Price (LMP) at
\texttt{SPPNORTH\_HUB} and were obtained from the publicly available SPP
Marketplace Portal.\footnote{\url{https://portal.spp.org/}} As in the German
case, the point forecasting models also use day-ahead forecasts of system load
and renewable generation. Load forecasts are taken from the SPP Mid-Term Load
Forecast (MTLF) reports, while the renewable forecast is constructed as the sum of
the available wind and solar forecasts from the Mid-Term Resource Forecast (MTRF)
reports. Both MTLF and MTRF are updated by SPP every hour. To reproduce a
day-ahead forecasting setting without using information that would not have
been available at the time of the market decision, for operating day $d$ we
use the forecast snapshot published at 09:00 Central Time on day $d-1$. If the 09:00 forecast is unavailable or incomplete, we use the most recent earlier available forecast vintage.

\begin{figure*}
    \centering
    \includegraphics[width=0.99\linewidth]{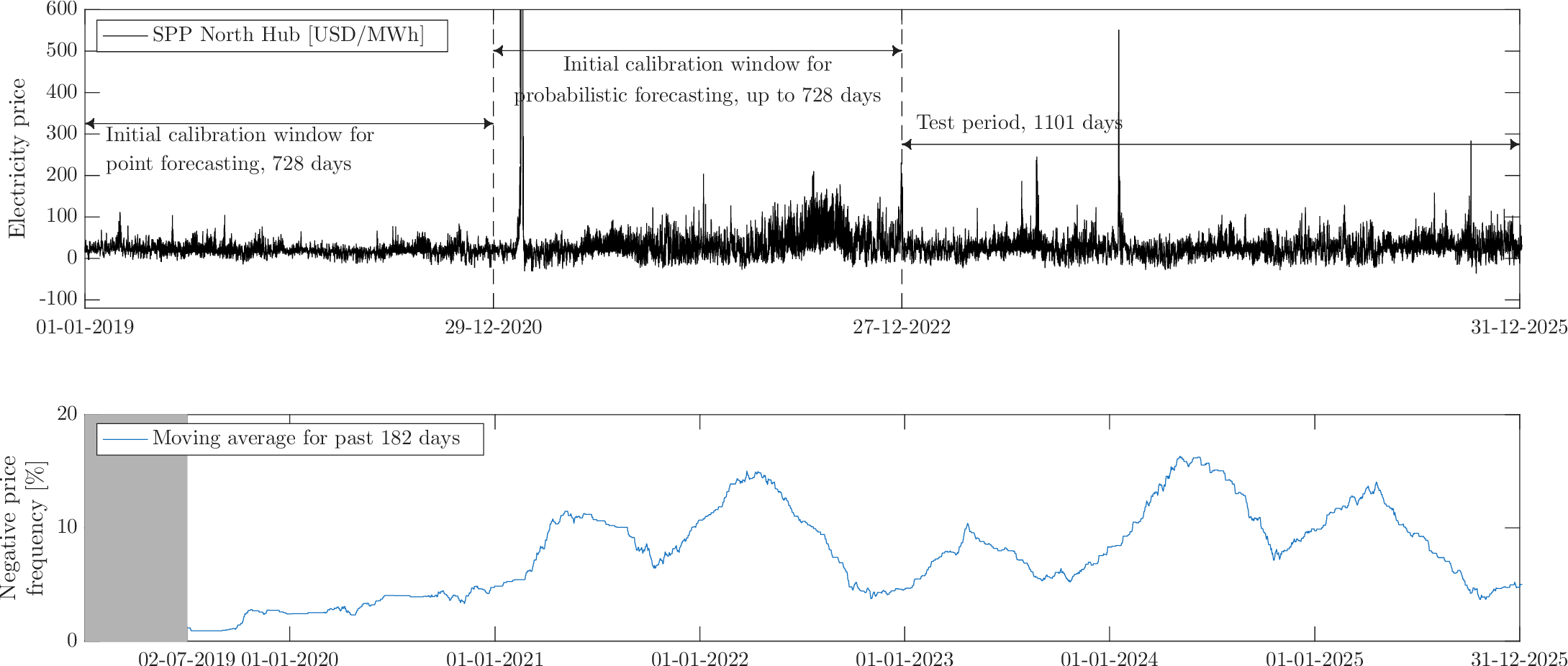}
    \caption{Hourly SPP North Hub day-ahead prices (top panel) and the rolling frequency of negative prices computed over a 182-day window (bottom panel). The first and second vertical dashed lines indicate the end of the initial calibration periods for point and probabilistic forecasting, respectively. For clarity, the price series is truncated at 600 USD/MWh; the maximum observed price was 4155 USD/MWh on 18-02-2021.}
    \label{fig:dataSPP}
\end{figure*}

\subsection{SPP results}
\label{app:spp_data}

Overall, the SPP results presented in Figures \ref{fig:crpsSPP}--\ref{fig:tail_quantile_inversionSPP} confirm the main findings of the German case study,
although the improvements in predictive accuracy are less pronounced. For the
full distribution and the tails, the differences between SI, SD, and SS are
generally small for ARX and more mixed for NARX, while LEAR shows a clearer
advantage of the sign-dependent specifications. The results conditional on
negative point forecasts are also less uniform, with NARX in particular showing
some atypical behaviour. Thus, the accuracy gains from sign dependence appear
to depend partly on the underlying point forecasting model.

In contrast, the results concerning forecast coherence are very similar to
those obtained for the German market. Independent estimation leads to frequent
quantile crossing and tail-quantile inversion, while SD-J and SS-J eliminate
both problems. SI-J continues to produce non-zero crossing frequencies and
tail-quantile inversions, confirming that joint estimation alone does not
resolve the structural problem of the sign-invariant specification.

An interesting difference between the two markets concerns the negative-price
regime. Negative prices occur considerably more frequently in SPP, but their
average negative price is much smaller (in absolute terms): approximately $-7.6$ USD/MWh in SPP compared
with about $-12.9$ EUR/MWh in Germany. The comparison therefore suggests that
both the frequency and the depth of negative prices matter. A higher frequency
increases the number of occasions on which the problem may arise, whereas more
extreme negative values amplify the mechanism described in Eq.~(2). This may
help explain why the coherence problem remains clearly visible in SPP while the
improvements in pinball score are less pronounced than in the German market.

\begin{figure*}[thb!]
    \centering
    \includegraphics[width=0.75\linewidth]{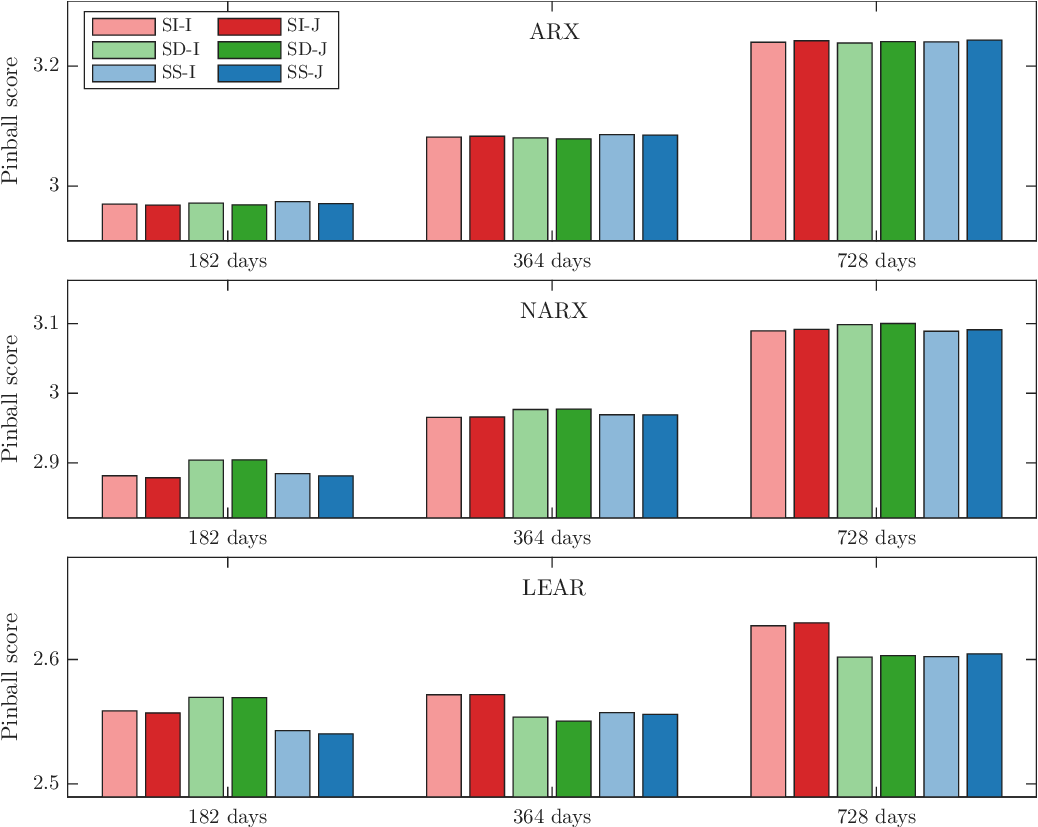}
    \caption{{Average pinball score for the considered QR-based postprocessing specifications for calibration window lengths of 182, 364, and 728 days, and point forecasting models ARX, NARX and LEAR. SI, SD, and SS denote the standard, sign-decomposed, and sign-specific variants, respectively, while -I and -J refer to the estimation schemes. Note: The y-axis scale differs across panels to enhance within-model comparability.}}
    \label{fig:crpsSPP}
\end{figure*}

\begin{figure*}[thb!]
    \centering
    \includegraphics[width=0.75\linewidth]{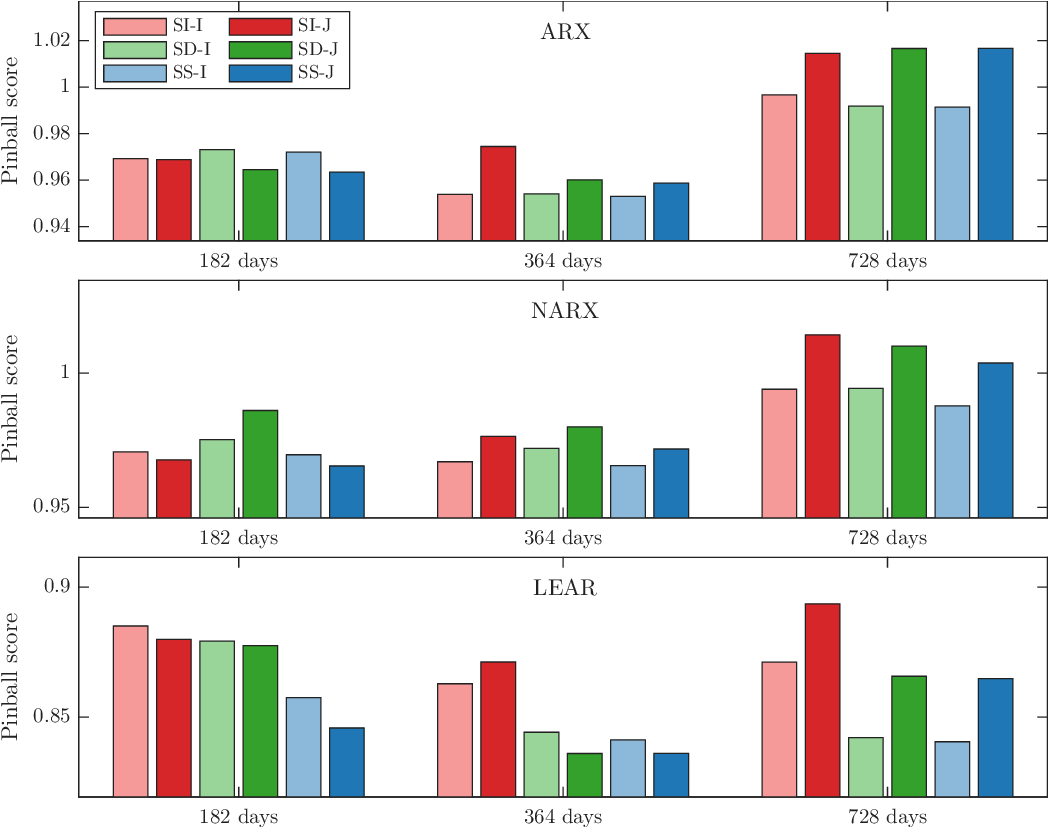}
    \caption{{Average pinball score computed over the 1st--5th and 95th--99th percentile forecasts. The layout and notation follow Figure~\ref{fig:crps}.}}
    \label{fig:crps_tailsSPP}
\end{figure*}

\begin{figure*}[thb!]
    \centering
    \includegraphics[width=0.75\linewidth]{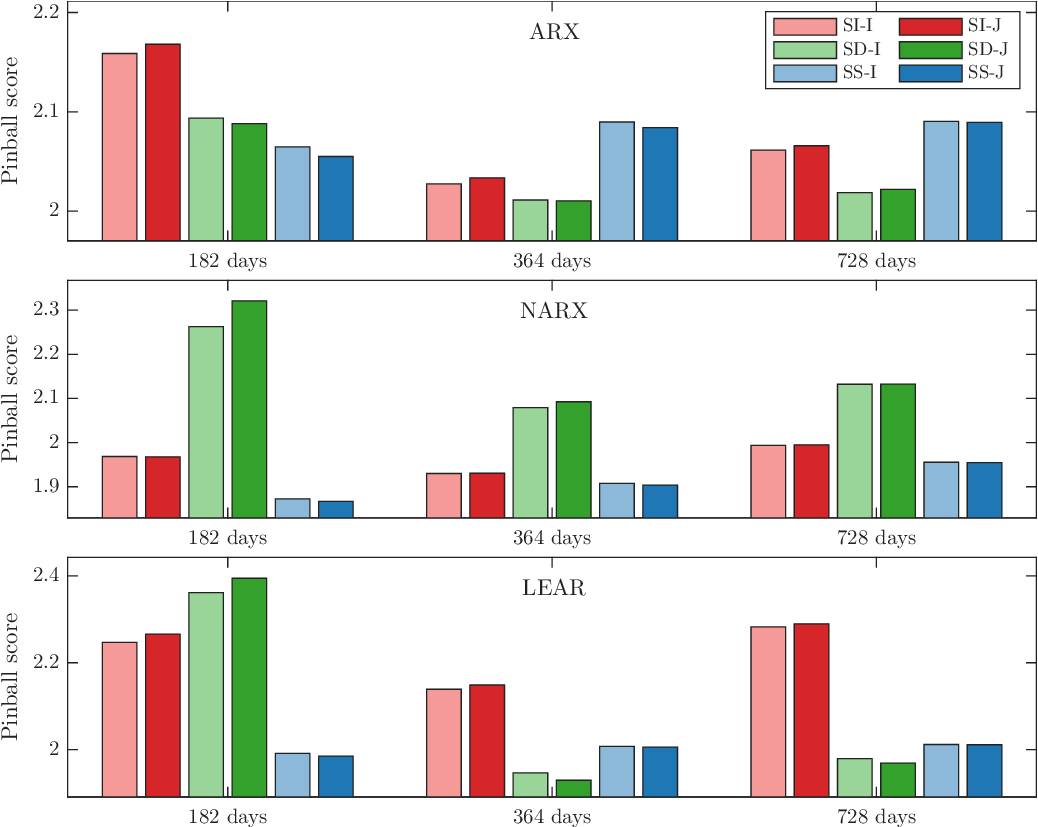}
    \caption{{Average pinball score computed only for observations with negative point forecasts. The layout and notation follow Figure~\ref{fig:crps}.}}
    \label{fig:crps_negativesSPP}
\end{figure*}

\begin{figure*}[thb!]
    \centering
    \includegraphics[width=0.75\linewidth]{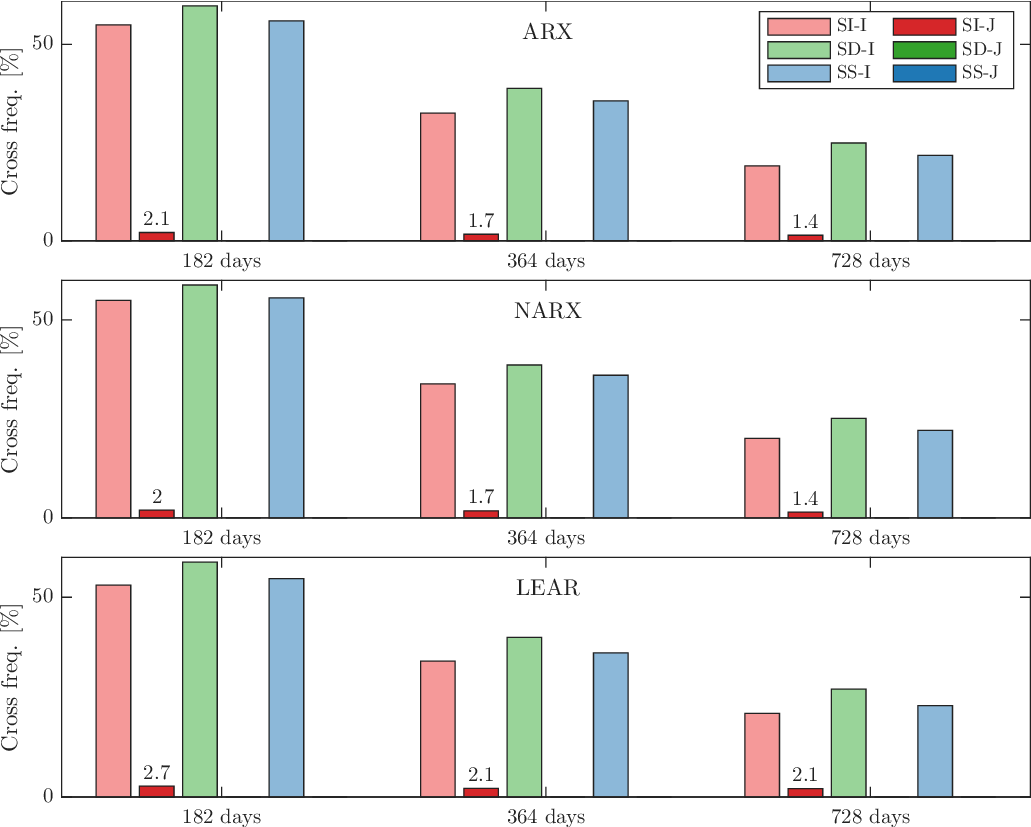}
    \caption{{Crossing frequency for the quantile forecasts. The layout and notation follow Figure~\ref{fig:crps}. Numerical labels are shown for SI-J to highlight that crossing persists under this standard specification even with joint estimation across quantiles.}}
    \label{fig:cross_incSPP}
\end{figure*}

\begin{figure*}[thb!]
    \centering
    \includegraphics[width=0.75\linewidth]{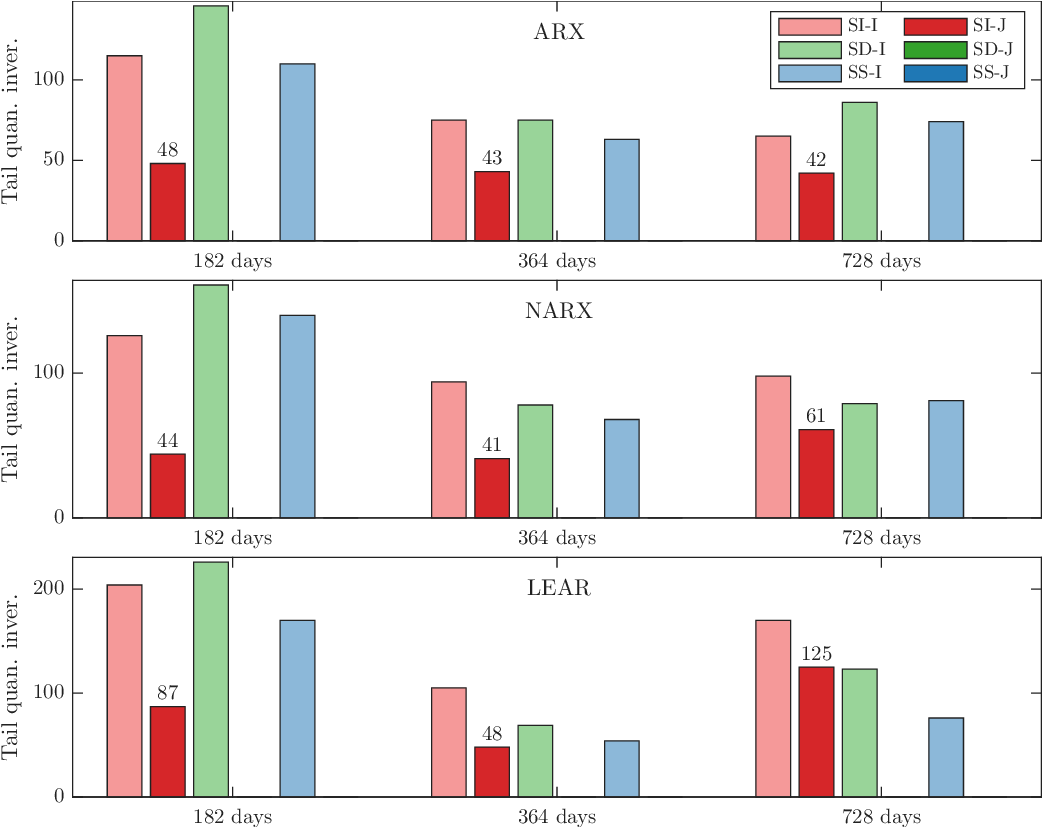}
    \caption{{Tail-quantile inversion count for the raw, unsorted quantile forecasts. The layout and notation follow Figure~\ref{fig:crps}. Numerical labels highlight the non-zero inversion counts for SI-J.}}
    \label{fig:tail_quantile_inversionSPP}
\end{figure*}
}

\end{document}